\documentclass[onecolumn,12pt,draftclsnofoot]{IEEEtran}
\usepackage[utf8]{inputenc}
\usepackage{pdflscape}
\usepackage{epsf}
\usepackage{graphicx}
\usepackage{epsfig,latexsym,amsmath,epsf,amssymb,amsfonts}
\usepackage{amsmath} 
\usepackage{amssymb}
\usepackage{amssymb}
\usepackage{placeins}
\usepackage{epsfig}
\usepackage{epstopdf}
\usepackage{color}
\usepackage[usenames,dvipsnames]{xcolor} 
\usepackage{algorithmicx}
\usepackage{amsthm} 
\usepackage{algorithm}
\usepackage{color}
\usepackage{algpseudocode}
\usepackage{algorithm}
 \usepackage{algorithmicx}
\usepackage{algpseudocode}
\usepackage{authblk}
\usepackage{multirow}
\usepackage{tabularx}
\usepackage[usenames,dvipsnames]{xcolor}				
\usepackage{setspace}
\usepackage{caption}
\usepackage{subcaption}
\usepackage{url}
\usepackage{subcaption}
\usepackage{lipsum}

\newtheorem{theorem}{Theorem} 
\newtheorem{lemma}[theorem]{Lemma}

\newtheorem{proposition}[theorem]{Proposition}
\newtheorem{corollary}[theorem]{Corollary}
\newtheorem{definition}{Definition}

\newtheorem{assumption}{Assumption} 
\newtheorem{data model}{Data Model}

\begin{document}
\title{A Game-Theoretic Framework for the Virtual Machines Migration Timing Problem}

\author[1]{Ahmed H. Anwar}
\author[1]{George Atia}
\author[2]{Mina Guirguis}

\affil[1]{\small Department of Electrical and Computer Engineering, University of Central Florida, Orlando, FL 32816}

\affil[2]{\small Department of Computer Science, Texas State University, San Marcos, TX 78666}
\affil[ ]{\small a.h.anwar@knights.ucf.edu, george.atia@ucf.edu, msg@txstate.edu}
\maketitle

\begin{abstract}

In a multi-tenant cloud, a number of Virtual Machines (VMs) are collocated on the same physical machine to optimize performance, power consumption and maximize profit. This, however, increases the risk of a malicious VM performing side-channel attacks and leaking sensitive information from neighboring VMs. To this end, this paper develops and analyzes a game-theoretic framework for the VM migration timing problem in which the cloud provider decides \emph{when} to migrate a VM to a different physical machine to reduce the risk of being compromised by a collocated malicious VM. The adversary decides the rate at which she launches new VMs to collocate with the victim VMs. Our formulation captures a data leakage model in which the cost incurred by the cloud provider depends on the duration of collocation with malicious VMs. It also captures costs incurred by the adversary in launching new VMs and by the defender in migrating VMs. We establish sufficient conditions for the existence of Nash equilibria for general cost functions, as well as for specific instantiations, and characterize the best response for both players. 
Furthermore, we extend our model to characterize its impact on the attacker's payoff when the cloud utilizes intrusion detection systems that detect side-channel attacks. Our theoretical findings are corroborated with extensive numerical results in various settings.

\end{abstract} 

\section{Introduction}
One of the main characteristics of the Cloud that allows scalable and cost-effective operation is multi-tenancy. Multi-tenancy is achieved through virtualization to enable cloud providers to host multiple virtual machines (VMs) on the same physical machine while providing isolation between them. Recent attacks, however, have been shown to bypass such isolation \cite{ren2012security}. A malicious VM collocating on the same physical machine with a victim VM can seek unauthorized access to sensitive and private data and/or intellectual property, or can render some of its computational functionality unusable. 

This has prompted cloud providers to develop various strategies for VM placement, migration and reconfiguration to mitigate some of these attacks. Moving target defense (MTD) strategies aim to dynamically shift the attack surface, making it more difficult for attackers to launch potent attacks \cite{csd}. When developing an MTD strategy, two main questions generally arise: \textit{which} targets should be moved? and \textit{when} should they be moved? The answer to these questions is highly-dependent on the context of the problem and the nature of the attack. For example, if an attacker contemplates inferring the underlying topology of the cloud, then the connectivity between machines is the target that should be changed over time. In a different setting, if the attacker is interested in cracking the system credentials that protect the users' databases, then the keys are the target that should be constantly reconfigured (i.e., moved). In this paper, we consider collocation attacks whereby an attacker can leak sensitive data from a targeted victim by running a VM on the same physical node (e.g., through launching a side-channel attack). Thus, for securing such system, VMs should be periodically migrated (i.e., moved to a different physical machine). This paper is primarily focused on the second question, that is, {\it when} to move the identified targets. 

In the MTD literature, this question is usually referred to as the timing problem of the MTD strategy. In this paper, we study this question using a game-theoretic framework seeking an understanding of the interplay of the actions of both the cloud provider (i.e., the defender) and the adversary. In our formulation, the adversary seeks to prolong the collocation time with the victim VMs to maximize information leakage. Since the adversary has no guarantees to be successfully collocated on the same physical machine with the victim -- since different cloud providers implement different placement algorithms according to different criteria that the attacker has no control over -- her best-effort would be to increase the number of VMs to launch (which is a cost metric we capture). The adversary can then check after being placed whether she had a successful collocation or not \cite{yarom2014flush+}. The cloud provider, on the other hand, seeks to migrate VMs between physical machines to minimize the collocation times between VMs. VM live migration, while efficient, is not free \cite{voorsluys2009cost} and thus the question so as to when to migrate is crucial in order to mitigate the collocation attack threats while not burdening the system with a large overhead that may not be justified.

\noindent {\bf Contributions:}  While VM migration strategies have been proposed as defense mechanisms against collocation attacks in various studies, such work focused on the VM assignment problem (mapping VMs to physical nodes) as a single player scheduling problem. In this paper, however, we consider the timing problem of the MTD as a game between the attacker and the cloud provider. Our work contributes to the theory of timing games \cite{blackwell1949noisy,radzik1996results}, which is largely unexplored in cloud computing settings. We leverage the results of the leakage model in the FlipIt game considered previously in \cite{bowers2012defending,van2013flipit,hu2015dynamic,zhang2014stealthy,pawlick2015flip,farhang2016flipleakage} to develop a novel formulation to study the VM collocation problem in an extended FlipIt game-theoretic framework. To the best of our knowledge, this is the first work to investigate the following aspects of the timing games.

\begin{itemize}
\item We provide a new game-theoretic formulation for the VM collocation timing problem.
\item Unlike \cite{han2013security,kamhoua2014game,kamhoua2015cyber}, we do not assume the defender has prior knowledge of the exact location of the attacker, thereby allowing for realistic threat and defense models. The defender has to migrate the VMs at the right time(s) to defend against malicious collocating users.
 
\item We analytically characterize the Nash equilibrium (NE) for the studied game model and derive sufficient existence conditions. 

\item We study the behavior of the adversary when the defender adopts an intrusion detection system (IDS). In this case, the adversary not only takes attack actions, but also decides when to stop her attack to reduce the risk of being detected.

\item We provide extensive numerical experiments to support our theoretical findings and compare our proposed defense policies against other defense policies. In our numerical evaluation, we consider several reward functions to reflect the severity of the attack and different degrees of information leakage.
\end{itemize}

This paper is organized as follows. In Section  \ref{sec:model}, we provide the system model and game formulation. In Section \ref{sec:analysis}, we provide theoretical analysis and establish existence conditions of NE for the formulated game. Our numerical results are presented in Section \ref{sec:results} and we conclude the paper in Section \ref{sec:conc}.

\section{Related work}\label{sec:relwork}
This work is at the intersection of two areas focused on securing cloud computing: Cross-VM side-channel attacks and mitigation, and the use of game theory in modeling the interplay between the cloud provider and the adversary. In this section we put our work in context within these two areas.

\subsection{Cross-VM side channel attacks and mitigation strategies}
Cloud security has received considerable attention recently \cite{ren2012security,singh2016survey}. Various studies have investigated the impact of cross-VM side-channel attacks \cite{ristenpart2009hey,suzaki2011memory,owens2011non,yarom2014flush+,zhang2012cross,liu2015last,irazoqui2014wait}. Users cryptographic keys have been shown to be vulnerable to exfiltration attacks when adversaries perform Prime+Probe attacks on the square-and-multiply implementation of GnuPG \cite{zhang2012cross}. The authors in \cite{yarom2014flush+,irazoqui2014wait,liu2015last} have shown that some side-channel attacks can extract cryptographic keys by exploiting the last-level shared caches of the memory. Other attacks have identified pages that a VM shares with its collocated neighboring VMs revealing information about the victim’s applications \cite{suzaki2011memory} and OS \cite{owens2011non}.

To combat cross-VM side-channel attacks, various approaches have been proposed at the hypervisor \cite{vattikonda2011eliminating,li2014stopwatch,zhang2012cross,raj2009resource,kim2012stealthmem}), the guest OS \cite{zhang2013duppel}, the hardware level \cite{wang2008novel,liu2014random}, and the application layer \cite{pattuk2014preventing}. These techniques, however, suffer from two fundamental limitations. First, they cannot be generalized to different types of side-channel attacks \cite{zhang2014cross}. Second, they require major changes to the hypervisor, OS, hardware, and applications \cite{moon2015nomad}. VM live migration, on the other hand, has been proposed as an effective mechanism to combat side-channel attacks \cite{voorsluys2009cost,shrivastava2011application}. The authors in \cite{zhang2016cloudradar} provided a detection mechanism known as \textit{CloudRadar} that works as a real-time side-channel attack detector based on monitoring hardware performance counters. The authors in \cite{zhang2011homealone} proposed another detection system that can differentiate between friendly and other malicious activities of neighboring tenants. The authors in \cite{qi2016generic} showed that by controlling the placement process, a defense mechanism can mitigate the effect of cross-VM attacks through reducing the co-run probability between users. The approach, however, is only effective in the case of time-sensitive attacks and when the number of assigned virtual CPUs is large.
Motivated by the Moving target defense (MTD) concept, the authors in \cite{sun2016selance} presented a migration engine in which VMs are migrated to balance the load between different nodes in the cloud. Although MTD is a well-known defense methodology, the authors in \cite{achleitner2017stealth} demonstrated that in certain scenarios the migrated VMs can be tracked by adversaries. Hence, they proposed a stealthy approach to migrate VMs that can hide them on the network. In \cite{penner2017combating}, the authors study an MTD migration strategy against an attacker solving a multi-armed bandit problem seeking to collocate VMs with high rewards.

\subsection{Cloud Security using Game-Theoretic Techniques}
The use of game theory has largely focused on the VM allocation problem in the presence of adversaries \cite{han2013security,kamhoua2014game,kwiat2015security,kamhoua2015cyber,prakash2015empirical}. A common assumption in such formulations is that the adversary is known which does not typically hold in practice. Additionally, existing formulations do not consider the timing question for the VM migration problem, which is a critical one for the cloud provider wishing to migrate VMs for security. 
A more practical leakage model was considered in \cite{laszka2013mitigation,johnson2015games}, based on the FlipIt game model.  
FlipIt is a two-player game in which a defender and an attacker compete over the control of a given resource which can only be held by one player at a time. A flip is an action taken by a player to gain control of the resource. The goal is to hold the resource for as long as possible with the least number of flips (i.e., flips are costly). Over time, the resource generates rewards for the player holding the resource. The state of the resource is obscured form each player until they ``flip". Several variants of the FlipIt game model were considered to study different security situations \cite{bowers2012defending,van2013flipit,pham2012we,feng2015stealthy,hu2015dynamic,zhang2014stealthy,pawlick2015flip,farhang2016flipleakage}. In \cite{van2013flipit}, the authors studied different strategies for each player and calculated dominant strategies and Nash equilibria. 
In \cite{pham2012we}, the game model was extended under the assumption that the players know the state of the resource before taking actions. In \cite{feng2015stealthy,hu2015dynamic} the game was extended to the case of a system where insiders can work in favor of external adversaries. The authors in \cite{zhang2014stealthy} considered the game with both players having limited budgets. Pawlick et al. investigated the game model with characteristics of signaling games \cite{pawlick2015flip}. In \cite{farhang2016flipleakage}, Farhang et al. studied a variant of the FlipIt game with an associated data leakage model in which the defender can partially eliminate the foothold of the attacker. The attacker exploits the system vulnerabilities that appear based on a periodic process. The authors assume that the attacker's strategy is fixed since she always starts to attack right after the defender takes his action. This, however, requires the attacker to fully observe the defender's strategy which we do not assume here.

In this work, we consider a significantly different and a realistic threat model that captures data leakage due to cross-VM side-channel attacks and develop defense strategies for identifying the best time(s) to migrate VMs. We do this through a game-theoretic framework in which the attacker only controls the attack rate and does not fully observe the defender's strategy. In addition, we assume that the attacker controls the probability of a successful attack by choosing the attack rate as opposed to the time to launch the attack.

\section{System Model}\label{sec:model}
\subsection{The cloud}

\begin{figure}
\centering
\includegraphics[width= 1\linewidth,height=0.1\textheight]{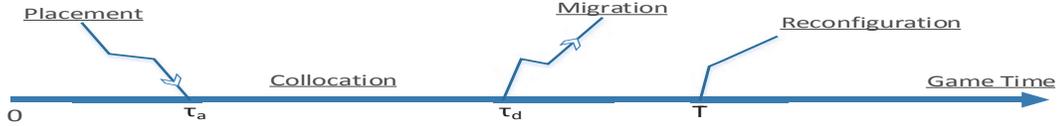}
\centering\caption{System model illustration}
\label{fig:system1}
\vspace{-0.6 cm}
\end{figure}

We model the cloud as a set of physical machines whereby each machine can host a number of VMs from different users.
The cloud provider uses a placement strategy to initially assign VMs to physical machines. The details of the placement strategy do not affect our analysis and we assume that the adversary (or any user) has no control over it. We assume the adversary is interested in targeting a set of victim VMs by collocating with them on the same physical machine. We study the interaction between the cloud provider (defender) and the adversary through a game-theoretic framework in which the rewards are time-dependent. In particular, the defender's strategy is to choose the time to re-assign VMs to different machines to defend against collocation attacks. The adversary, on the other hand, chooses an attack rate to launch more VMs to increase her collocation duration to maximize information leakage from her victims as described in Fig.~\ref{fig:system1}.
We define the game next.

\subsection{The Game}\label{sec:game}

A game is defined as a tuple ${\Gamma}(\mathcal{P}, \mathcal{A},\mathcal{U})$, where

\begin{itemize}
\item $\mathcal{P}$ is the set of players. Here, $\mathcal{P}=\left \{  1,2\right \}$, denoting the defender (player 1) and the adversary (player 2).
\item $\mathcal{A} =   \mathcal{A}_d \times \mathcal{A}_a $ is the action space for the defender and adversary. 
\item $\mathcal{U}$ $=\left \{ {u}_d , {u}_a\right \}$ is the reward function, $\mathcal{U}: \mathcal{A} \rightarrow \mathbb{R}^2$.
\end{itemize}

\subsubsection{Defender's action space}
Since we are investigating the timing factor, the cloud provider (referred to as the system defender) is assumed to control the re-allocation period. Let $\tau_d \in \mathcal{A}_d$ denote the time instant at which the defender migrates a running VM to a new physical node, such that $\mathcal{A}_d = [\tau_{\min}, T]$, where $T$ is a system parameter at which the credentials are reset and $\tau_{\min}$ the smallest reconfiguration time. Since we assume a leakage model, at time $T$ when the system credentials are reset, the attacker can no longer benefit from the side-channel attack. Therefore, the whole game will be reset every $T$. The defender seeks to optimize the value of $\tau_d$ to minimize chances for information leakage and avoid loading the system with unnecessary migrations. Thus, the defender's goal is to optimize the tradeoff between security and stability. In particular, 
a smaller $\tau_d$ ensures the system is more secure since the co-residency times between any two VMs will be small. However, the system's overhead increases due to the frequent migration of the VMs between the physical nodes. On the other hand, a larger $\tau_d$ leads to a more stable system. However, the co-residency times between VMs on the same node will be large making the system more susceptible to information leakage through collocation attacks.

\subsubsection{Attacker's action space}

Here, we assume that the attacker does not know the system placement algorithms, hence only tries to increase her co-residency chances via increasing the number of requests submitted to the cloud provider. Let $\lambda_a \in \mathcal{A}_a$ denote the rate of requests (rate of attack) submitted to the cloud, where $\mathcal{A}_a = [\lambda_{\min},\lambda_{\max}]$ is an interval of non-negative attack rates. The game is assumed to start at time $t=0$, and let $\tau_a$ denote the actual time at which the attacker successfully collocates with her targeted victim. Hence, $\tau_a > 0 $ is a non-negative random variable with a probability density function (pdf) $f_a(.;\lambda_a)$ parametrized by $\lambda_a$. Since the attacker pays a cost for each submitted job, she needs to optimize over the attack rate $\lambda_a$. Hence, the attacker's tradeoff can be summarized as follows.
When $\lambda_a$ is very small, it is less probable for the attacker to successfully co-reside with her victim and in turn leak any information before it is migrated. When $\lambda_a$ is very large, the attacker increases her chances of successful collocation at the expense of a higher attack cost.
Therefore, the pdf $f_a$ should be such that $f_a(\tau_a; \lambda_{a_1})$ yields a higher probability of early collocation than $f_a(\tau_a; \lambda_{a_2})$, when $\lambda_{a_1} > \lambda_{a_2}$. Mathematically, this requirement is expressed in the following assumption.
\begin{assumption}\label{assumptio:fa}
$F_{a}(t; \lambda_{a_1}) \geq F_{a}(t; \lambda_{a_2}) $ for $\lambda_{a_1} \geq \lambda_{a_2}$, where $F_a(t;\lambda_a) := \Pr (\tau_a \leq t)$ denotes the cumulative distribution function (CDF) of the collocation time.
\end{assumption}
If $\lambda_{\min} = 0$, then the attacker can choose to back off (i.e., not attack). In such case, $f_a(\tau_a;0)$ is a degenerate deterministic distribution such that $F_a(T;0) = 0$ since the probability of collocation is 0. Since the game ends at $\tau_d$ then repeated, we consider the reward per unit time. 
We focus only on the timing factor of the problem, and the mapping of VMs to physical nodes is carried out through the placement engine. 
Any newly arriving VM or existing VMs that are being migrated can be passed to the placement engine for allocation, hence no system hardware modification is required. Next, we define the players' reward (payoff) functions. We assume a nonzero-sum two-person game.

\subsubsection{Attacker's reward} 
Once the attacker is successfully placed on the same node where the victim VM resides, she immediately starts accumulating rewards by leaking information. Let $G(\tau_d,\tau_a) $ denote the reward accumulated by the attacker. 
\begin{assumption}\label{assumption:G}
$G(\tau_d,\tau_a) = G(t)$ is  a stationary function and monotonically non-decreasing in the collocation duration $t$, where $t = \tau_d - \tau_a$.
\end{assumption}
Stationarity signifies that the attacker's accumulated reward depends on the collocation and migration times only through their difference, i.e., the duration of collocation. 
The accumulated reward is assumed to be zero if $\tau_a \geq \tau_d $. The attacker incurs a cost $C_a$ for launching this attack. Hence, the total cost is scaled by the rate of attack $\lambda_a$. Therefore, the attacker's payoff is given by \begin{equation}\label{eqn:attacker_rew}
\tilde{u}_a(\tau_d,\tau_a,\lambda_a) =  \frac{1}{\tau_d} \left [   G(\tau_d,\tau_a)  \cdot  \mathbf{1} _{\{\tau_a <\tau_d\}} - \lambda_a C_a \right ]
\end{equation}
where $\mathbf{1}_{\{\}}$ is an indicator function, and the tilde notation signifies the payoff for a given realization of $\tau_a$, which is a random variable. Hence, the expected payoff is
\begin{equation}\label{eqn:attacker_cond_rew}
u_a(\tau_d,\lambda_a) =  {\int_{0}^{\infty}\tilde{u}_a(\tau_d,\tau_a,\lambda_a)f_a(\tau_a,\lambda_a)\:d\tau_a}\:.
\end{equation}
\subsubsection{Defender's reward} 

The defender, on the other hand, incurs a loss due to the collocation of a victim VM with the attacker equal in magnitude to the gain of the attacker. 
In addition, the defender pays a cost per migration, which increases the system overhead and overloads the placement engine. The cost of migration is denoted by $C_d$. Accordingly, the defender's payoff can be written as
\begin{equation}\label{eqn:defender_rew}
\tilde{u}_d(\tau_d,\tau_a,\lambda_a) =  \frac{1}{\tau_d} \left [-G(\tau_d,\tau_a)  \cdot \mathbf{1} _{\{\tau_a <\tau_d\}}  -   C_d \right ]\:.
\end{equation}
Averaging over $\tau_a$, the expected payoff for the defender can be calculated as
\begin{equation}\label{eqn:defender_cond_rew}
u_d(\tau_d,\lambda_a) =  {\int_{0}^{\infty}\tilde{u}_d(\tau_d,\tau_a,\lambda_a)f_a(\tau_a,\lambda_a)d\tau_a}\:.
\end{equation}
The probability of successful collocation (i.e, $\tau_a <\tau_d$) is computed as
\begin{equation}
P(\tau_a <\tau_d) = \int_{0}^{\tau_d} f_a(\tau_a,\lambda_a) d\tau_a\:.
\end{equation}

\section{Theoretical Analysis}\label{sec:analysis}

In this section, we derive sufficient conditions for the existence of Nash equilibria for the formulated game. Existence of Nash equilibria depends on the properties of the payoff functions. 
First, we derive existence conditions for a general accumulated reward function $G(\tau_d,\tau_a)$ and pdf of the collocation time $f_a(\tau_a;\lambda_a)$, then we provide analysis for a specific instantiation of the payoff functions. 
We also characterize the best response curves for both players and derive conditions for Nash equilibrium strategies if they exist. 
First, we restate a general theorem from \cite{bacsar1998dynamic} that provides sufficient conditions for $N$-person nonzero-sum games to admit a pure strategy Nash equilibrium.  

\begin{theorem}\label{theorem1} \cite{bacsar1998dynamic}
For each player $i$ in the set $\mathcal{N}$ of $N$ players, let the action space $U_i$ of player $i$ be a closed, bounded and convex subset of a finite-dimensional Euclidean space, and the cost functional $J_i:U_1 \times \cdots \times U_N \to \mathbf{R}$  be jointly continuous in all its arguments and strictly convex in $u_i\in U_i$, for every $u_j \in U_j, j \in \mathcal{N}, j \neq i$. Then, the associated $N$-person nonzero-sum game admits a Nash equilibrium in pure strategy.
\end{theorem}

\subsection{General reward functions}
\label{subsec:generic}
For the general payoff formulation described in equations (\ref{eqn:attacker_cond_rew})
and (\ref{eqn:defender_cond_rew}), the following lemma proved in the appendix establishes sufficient conditions for the concavity of the payoff functions.  

\begin{lemma}\label{theorem: general}
\textit{For the 2-person nonzero-sum game defined in Section \ref{sec:game} with payoff functions defined in equations (\ref{eqn:attacker_cond_rew}) and (\ref{eqn:defender_cond_rew}) under Assumptions \ref{assumptio:fa} and \ref{assumption:G}, if $f_a(\tau_a;\lambda_a)$ is strictly concave in $\lambda_a \in \mathcal{A}_a$, then  $u_a(\tau_d,\lambda_a)$ is strictly concave in $\lambda_a$ for any $\tau_d \in \mathcal{A}_d$, and if $\frac{G(\tau_d, \tau_a)}{\tau_d} $ is convex in $\tau_d \in \mathcal{A}_d$, then $u_d(\tau_d,\lambda_a)$ is strictly concave in $\tau_d$ for any $\lambda_a \in \mathcal{A}_a$.}
\end{lemma}

Therefore, we can readily state sufficient conditions for our game to admit a pure strategy Nash equilibrium. 

\begin{theorem}\label{theorem_NE}
The 2-person nonzero-sum game defined in Section \ref{sec:game} under Assumptions \ref{assumptio:fa} and \ref{assumption:G} with the payoff functions in (\ref{eqn:attacker_cond_rew}) and (\ref{eqn:defender_cond_rew}) admits a  Nash equilibrium in pure strategy if $f_a(\tau_a;\lambda_a)$ is continuous and strictly concave in $\lambda_a \in \mathcal{A}_a$, and $\frac{G(\tau_d, \tau_a)}{\tau_d} $ is convex and $G$ is continuous in $\tau_d \in \mathcal{A}_d$ . 
\end{theorem}
%
The proof of Theorem \ref{theorem_NE} follows directly from Lemma \ref{theorem: general}, which establishes strict concavity of the payoff functions under the conditions in the statement of the theorem, and Theorem \ref{theorem1} from \cite{bacsar1998dynamic}.

\begin{proposition}\label{prop}
For the game defined in Section \ref{sec:game} with $\lambda_{\min} = 0$, there exists an equilibrium in which the attacker backs off (i.e., does not attack) and the defender does not migrate if the reward function $G(t)$ satisfies
\begin{equation}\label{eqn:equ_condn}
 \mathbb{E}_{\lambda_a}\left [ G(T-\tau_a) \right ] \leq \lambda_a C_a, 
\end{equation}
for every $\lambda_a \in \mathcal{A}_a$, where $\mathbb{E}_{\lambda_a}[.]$ denotes the expectation w.r.t. the measure induced by $f(.;\lambda_a)$. 
\end{proposition}

\begin{proof}
 If the attacker backs off, i.e., chooses $\lambda_a =\lambda_{\min} = 0$, then the defender's payoff in (\ref{eqn:defender_cond_rew}) becomes
\begin{equation*}
u_d(\tau_d, 0) = \frac{-C_d}{\tau_d},
\end{equation*}
which attains its maximum at $\tau_d = T$ for any $C_d > 0$. Hence, the defender's best response is to not migrate over the game interval. Also, if the condition (\ref{eqn:equ_condn}) in the statement of Proposition \ref{prop} is satisfied, then the attacker's best response to the defender's action $\tau_d = T$ is $\lambda_a = 0$. To see that note that if 
\begin{equation*}
\mathbb{E}_{\lambda_a}\left [ G(T,\tau_a) \right ]  = \int_0^{\infty} G(T,\tau_a) f_a(\tau_a;\lambda_a) d \tau_a \leq \lambda_a C_a .
\end{equation*} 
then,  
\begin{equation*}
\int_0^{\tau_d} G(\tau_d,\tau_a) f_a(\tau_a;\lambda_a) d \tau_a \leq   \lambda_a C_a
\end{equation*}
since $G(t)$ is monotonically non-decreasing in $t$ per Assumption \ref{assumption:G}.
Recalling the attacker's payoff function in (\ref{eqn:attacker_cond_rew}), the attacker's decision to back off is at least as good as launching an attack at an alternative non-vanishing rate since the cost of the attack upper bounds the leakage reward for any $\lambda_a \neq 0$.

\end{proof}

\begin{definition}
\label{def:br}
In an N-person nonzero sum game, let $u_i(a_1,\ldots, a_i,\ldots, a_N)$ be the reward function of player $i$. For each player $i \in \{ 1, \ldots,N \}$, assume that the maximum reward of $u_i$ with respect to $a_i \in \mathcal{A}_i$ can be attained for any players' action profile $a_{-i} \in \mathcal{A}_{-i}$, where $a_{-i} :=\left \{ a_1, \ldots,a_{i-1},a_{i+1}, \ldots, a_N \right \}$ and $\mathcal{A}_{-1} \equiv \mathcal{A}_{1} \times \ldots \mathcal{A}_{i-1}\times \mathcal{A}_{i+1}  \times \ldots \times \mathcal{A}_{N}  $. Then, the set $R_i(a_{-i}) \subset \mathcal{A}_i $ defined by
\[ R_i(a_{-i}) = \left \{ \zeta \in  \mathcal{A}_i : u_i( \zeta , a_{-i}) \geq u_i( a_i , a_{-i}), ~\forall a_i \in  \mathcal{A}_i \right \},
\]
is called the optimal (or best) response of player $i$. If $R_i$ is a singleton for every $a_{-i} \in \mathcal{A}_{-i} $, then it is called the reaction curve \cite{bacsar1998dynamic}.
\end{definition}
Accordingly, it follows from the definition of a Nash equilibrium (in that no player can gain by a unilateral change of strategy if the strategies of the other players remain unchanged) that the intersection points of the best responses are Nash equilibria. In the following theorem, we characterize the best response for both players.

\begin{theorem}\label{theorem_BR}
For the 2-person nonzero sum game defined in Section \ref{sec:game}, if the attacker's payoff function in (\ref{eqn:attacker_cond_rew}) is strictly concave in $\lambda_a$, then the attacker's best response $\lambda_a^*$ to any defense strategy can be described as
\begin{itemize}
\item $\lambda^*_a = \lambda_{\max}$, if $\frac{\partial u_a}{\partial \lambda_a} > 0, ~ \forall ~ \lambda_a \in \mathcal{A}_a$ 
\item $\lambda^*_a = \lambda_{\min}$
, if $\frac{\partial u_a}{\partial \lambda_a} < 0, ~ \forall ~ \lambda_a \in \mathcal{A}_a$
\item $\lambda^*_a   \in \left \{ \lambda_a  \mid  \int_0^{\tau_d} G(\tau_d,\tau_a) \frac{\partial f_a(\tau_a;\lambda_a)}{\partial \lambda_a} d\tau_a = C_a   \right \},$ if $\frac{\partial u_a}{\partial \lambda_a} = 0$, for any  $\lambda_a \in \mathcal{A}_a$.
\end{itemize} 
\noindent
Also, if the defender's payoff function in (\ref{eqn:defender_cond_rew}) is strictly concave in $\tau_d$, then the best response $\tau_d^*$ can be described as
\begin{itemize}
\item $\tau^*_d = T $, if $\frac{\partial u_d}{\partial \tau_d} > 0, ~ \forall ~ \tau_d \in \mathcal{A}_d$
\item $\tau^*_d =  {\tau_{\min}}$
, if $\frac{\partial u_d}{\partial \tau_d} < 0, ~ \forall ~ \tau_d \in \mathcal{A}_d$
\item $\tau^*_d  \in \left \{ \tau_d  \mid  \int_0^{\tau_d}\left ( \tau_d  \frac{\partial G}{\partial \tau_d} - G \right ) f_a(\tau_a;\lambda_a)  d\tau_a = C_d   \right \},$ if $\frac{\partial u_d}{\partial \tau_d} = 0$, for any  $\tau_d \in \mathcal{A}_d$.
\end{itemize}
\end{theorem}
\begin{proof}
Given the concavity of the payoff function $u_a$ in $\lambda_a \in \mathcal{A}_a$, the derivative $\frac{\partial u_a}{\partial \lambda_a}$ is monotone. Hence,
there exist three possibilities for the behavior of $u_a$: if $\frac{\partial u_a}{\partial \lambda_a} > 0$, then $u_a$ is strictly increasing in $\lambda_a$ for all $\lambda_a \in \mathcal{A}_a$, thus the payoff is maximized by $\lambda_a^* = \lambda_{\max}$. If $\frac{\partial u_a}{\partial \lambda_a} < 0, ~ \forall \lambda_a ~ \in \mathcal{A}_a$, then $u_a$ is strictly decreasing in $\lambda_a$ for all $\lambda_a \in \mathcal{A}_a$, thus the payoff is maximum at $\lambda_a^* = \lambda_{\min}$. Otherwise, $u_a$ attains its maximum when $\frac{\partial u_a}{\partial \lambda_a} = 0$, hence the best response $\lambda_a^*$ belongs to the set $\Lambda_a = \left \{ \lambda_a \mid  \int_0^{\tau_d}\frac{\partial f_a}{\partial \lambda_a} G(\tau_d,\tau_a) d\tau_a = C_a \right \}$ at which $\frac{\partial u_a}{\partial \lambda_a} = 0$. 
The second part of Theorem \ref{theorem_BR}, which characterizes the defender's best response can be proven similarly.
\end{proof}
Next, we study the effect of the attack cost $C_a$ and the moving cost $C_d$ and state bounds on the costs beyond which no player is interested in the game. When the cost $C_a$ exceeds a certain threshold, 
the cost of the attack dominates the attacker's tradeoff, i.e., the attacker is better off backing off over attempting to leak information. 
Similarly, if $C_d$ is too high, the defender incurs a cost for migration that exceeds any benefit he would get at any migration rate.

In the following lemma, we derive a lower bound on the attack cost $C_a$ beyond which the attacker is always better off attacking with the minimum rate $\lambda_{\min}$. If $\lambda_{\min} = 0$, then the attacker will back off.

\begin{lemma}\label{lemma:att_monotone}
For the two person nonzero-sum game $\Gamma$ defined in Section \ref{sec:game}, if $f_a$ is strictly concave in $\lambda_a \in \mathcal{A}_a$, and $C_a > \int_0^{\tau_d}G f'_a(\lambda_{\min} )d \tau_a $, where $ f'_a(\lambda_{\min} ) = \frac{\partial f_a}{\partial \lambda_a} |_{\lambda_a = \lambda_{\min}}$, then the attacker's best response to any defense strategy $\tau_d$ is to attack at the minimum permissible rate $\lambda_{\min}$.
\end{lemma}

\begin{proof}
{We argue that under the condition stated in the lemma, the attacker's payoff is monotonically decreasing in $\lambda_a$. Hence, $\lambda_a^* = \lambda_{\min}$ is the attacker's best response to any $\tau_d$. To show that $\lambda_{\min}$ is the unique best response, assume for contradiction that there exists $\lambda^* =\lambda_1$ such that $ \lambda_1 \ne \lambda_{\min}$. 
If $C_a > \int_0^{\tau_d}G f'_a(\lambda_{\min} )\:d\tau_a $, then $u_a$ is monotonically decreasing, therefore $u_a(\lambda_{\min}) > u_a(\lambda_{1})$ since $ \lambda_1 > \lambda_{\min} $. Hence, $\lambda_1$ is not in the best response set.
}
The details of the proof are deferred to the Appendix.
\end{proof}
Similarly, the following lemma establishes a lower bound on the migration cost $C_d$ of the defender, beyond which it is more advantageous not to migrate before the system reconfiguration cycle $T$. 
\begin{lemma}\label{lemma:def_monotone}
For the two person nonzero-sum game $\Gamma$ defined in Section \ref{sec:game}, if $\frac{G}{\tau_d}$ is strictly convex and $G$ is continuous
in $\tau_d \in \mathcal{A}_d$, and $C_d >  T^2 \:\mathbb{E} \left [ {\tilde G}(T) \right ]$, then the action of not migrating any VM before $T$ is the defender's unique best response regardless of the attacker's strategy $\lambda_a$, where $\mathbb{E[.]}$ is the expectation with respect to $f(\tau_a;\lambda_a)$ and $ \tilde G (T) = \frac{d }{d \tau_d} \left ( \frac{G}{\tau_d} \right )|_{\tau_d = T}$. 
\end{lemma}
\begin{proof}
By an argument similar to the proof of Lemma \ref{lemma:att_monotone}, under the condition in the statement of the lemma, the defender's payoff is monotonically increasing in $\tau_d$. Hence, $T\in R_1(\lambda_a)$ for any $\lambda_a$. Establishing the uniqueness of $T$ as a best response action follows the same argument used in the proof of Lemma \ref{lemma:att_monotone}. The details are deferred to the Appendix.
\end{proof}

\subsection{Specific instantiation analysis}\label{subsec:special}
In Section \ref{subsec:generic}, we provided conditions for the existence of an equilibrium for generic reward functions. The conditions imposed were the strict concavity of $f_a$ in addition to the non-negativity, monotonicity and stationarity of $G$ (stationarity in that the accumulated reward depends on the collocation and migration times only through their difference, i.e., the duration of collocation). In this section, we study existence conditions for equilibrium and characterize the best response sets of both players for specific choices of the reward function $G$ and the collocation pdf $f_a(\tau_a;\lambda_a)$. Specifically, we provide an analysis for the case where $G(t)$ increases linearly in the collocation duration $t$. 
Hence, we analyze the formulated timing game for the following choice of $G$, 
 
\begin{equation}\label{eqn:linear_g_of_t}
G(\tau_d , \tau_a) =  
  \begin{cases}
  \alpha \left( \tau_d - \tau_a\right)  , &  \tau_a \leq \tau_d \leq T \\
    0, & \text{otherwise. }   
\end{cases}
\end{equation}

In Section \ref{subsec:G_scalings}, we provide numerical results on the best response for other (non-linear) functions, including when $G$ scales quadratically in $t$. Without loss of generality, we always consider $\alpha = 1$. 
The case $\alpha \neq 1$ corresponds to the case $\alpha = 1$ with the migration cost $C_d$ replaced by $\frac{C_d}{\alpha}$. 

Since the attacker controls the rate of attacks $\lambda_a$, in our numerical evaluation we consider an exponential pdf $f_a$ for the collocation time, i.e., 
\begin{equation}\label{eqn:f_a}
f_a(\tau_a;\lambda_a) = \lambda_a e^{-\lambda_a \tau_a}~, ~~~   \tau_a \geq 0.
\end{equation}
\noindent
This choice of $f_a(.;\lambda_a)$ is motivated by the interpretation of $\lambda_a$ as the rate of attacks launched by the adversary where $\lambda_a \in \mathcal{A}_a$. 

Next, we derive sufficient conditions for the existence of a Nash equilibrium for the choice of functions in (\ref{eqn:linear_g_of_t}) and (\ref{eqn:f_a}).

\begin{figure}
\centering
\includegraphics[width=0.66\linewidth,height=0.33\textheight]{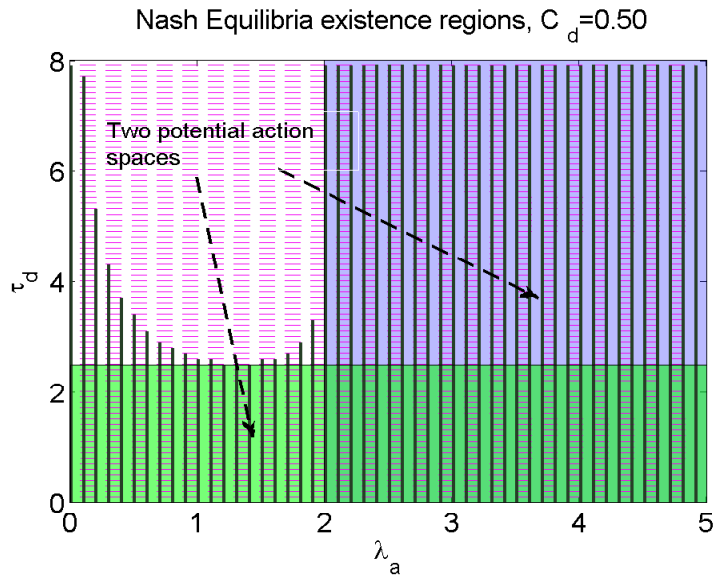}
\centering\caption{Games with different action spaces satisfying the existence condition for NE.}
\label{fig:action_spaces}
\end{figure}

\begin{theorem}\label{theorem2}
Consider the 2-person nonzero-sum game defined in Section \ref{sec:game} with $G(t)$ and $f_a(\tau_a;\lambda_a)$ defined in (\ref{eqn:linear_g_of_t}) and (\ref{eqn:f_a}).  
If $ 1-\lambda_aC_d < (1+\lambda_a \tau_d + \frac{\lambda^2_a \tau^2_d}{2})e^{-\lambda_a \tau_d}, \forall (\tau_d,\lambda_a) \in \mathcal{A}$, then the game admits a pure strategy Nash equilibrium.
\end{theorem}
The proof of Theorem \ref{theorem2} provided in the appendix rests upon establishing sufficient conditions for the strict concavity for $u_a$ and $u_d$, which translate into existence of a Nash equilibrium in pure strategy from \cite[Theorem 1]{bacsar1998dynamic}. In Fig. \ref{fig:action_spaces}, the region of intersection of the vertically and horizontally hashed areas represents the region of concavity of $u_d$ and $u_a$ in $\tau_d$ and $\lambda_a$, respectively, for all actions of the other player for $C_d = 0.5$. 
The figure also shows two different games with their corresponding action spaces satisfying the existence condition of Nash equilibrium of Theorem \ref{theorem2}. Later in Section \ref{sec:results}, we verify that games played on these action spaces admit a Nash equilibrium in pure strategies. 
The next results follow directly from Theorem \ref{theorem2}. 

\begin{corollary}
\label{cor:NE}
The 2-person nonzero-sum game $\Gamma$ defined in Section \ref{sec:game} with $\mathcal{A}_d =\left [ \tau_{\min},T \right ]$ and $\mathcal{A}_a = \left [ 1/C_d ,\lambda_{\max} \right ]$ admits a Nash equilibrium in pure strategies for any $T > 0$, $\tau_{\min} < T$ and $\lambda_{\max} > 1/C_d$.

\end{corollary}
\begin{corollary}
\label{cor:NE2}
The 2-person nonzero-sum game $\Gamma$ defined in Section \ref{sec:game} with $\mathcal{A}_d =\left [ \tau_{\min}, T \right ]$ and $\mathcal{A}_a = \left [ \lambda_{\min} ,\lambda_{\max} \right ]$ admits a Nash equilibrium in pure strategies for any $T \leq 5C_d$, $ \tau_{\min} > 0$ and $\lambda_{\min}, \lambda_{\max} \geq 0$.
\end{corollary}
The proof of Corollary \ref{cor:NE2} follows from the monotonicity of the RHS of the inequality in the statement of Theorem \ref{theorem2} in $\tau_d$ and the bound $1-x/5 < (1+x+x^2/2)e^{-x}$ for $x > 0$.

To characterize Nash equilibria for both players, we start off by characterizing the best response set for each player in the following lemma whose proof follows the same argument used in the proof of Theorem \ref{theorem_BR}. 

\begin{lemma}\label{lemma3} For the 2-person game defined in Section \ref{sec:game} with the reward function $G(t)$ and the probability density function $f_a(\tau_a;\lambda_a)$ defined in (\ref{eqn:linear_g_of_t}) and (\ref{eqn:f_a}), the attacker's best response pure strategy is characterized as
\begin{itemize}
\item $\lambda_a^* = \lambda_{\max}$\:, if 
~$
 e^{-\lambda_a \tau_d}\left ( (1-C_a \lambda_a^2) e^{\lambda_a \tau_d} - \lambda_a \tau_d - 1  \right ) > 0
$ 
\item $\lambda_a^* = \lambda_{\min}$\:, if ~ 
$
 e^{-\lambda_a \tau_d}\left ( (1-C_a \lambda_a^2) e^{\lambda_a \tau_d} - \lambda_a \tau_d - 1  \right ) < 0 $

\item $\lambda_a^* \in \left \{ \lambda_a \mid    e^{-\lambda_a \tau_d}\left ( (1-C_a \lambda_a^2) e^{\lambda_a \tau_d} - \lambda_a \tau_d - 1  \right ) = 0 \right \}$\:, otherwise,
\end{itemize}
for any action $\tau_d$ by the defender.

The best response strategy for the defender can be characterized as  
\begin{itemize}
\item $\tau^*_d = T $\:, ~ if $
 e^{-\lambda_a \tau_d}\left ( (C_d \lambda_a -1) e^{\lambda_a \tau_d} + \lambda_a \tau_d + 1  \right ) > 0$
\item $\tau^*_d =  {\tau_{\min}}$
, ~if $
 e^{-\lambda_a \tau_d}\left ( (C_d \lambda_a -1) e^{\lambda_a \tau_d} + \lambda_a \tau_d + 1  \right ) < 0$
\item $\tau^*_d \in \left \{ 
 e^{-\lambda_a \tau_d}\left ( (C_d \lambda_a -1) e^{\lambda_a \tau_d} + \lambda_a \tau_d + 1  \right ) = 0   \right \},$ otherwise,
\end{itemize}

for any action $\lambda_a$ by the attacker.
\end{lemma}

The following two theorems whose proof is provided in Appendix B establish bounds on both the attack cost $C_a$ and the migration cost $C_d$ beyond which the players' best response strategies are on the boundaries of their action intervals.

\begin{theorem}\label{theorem5}
For the two person nonzero-sum game defined in Section \ref{sec:game} with the reward function in (\ref{eqn:linear_g_of_t}) and the exponentially distributed collocation time $\tau_a$ in (\ref{eqn:f_a}), 
if \[C_a > \frac{1-(1+\lambda_{\max}\tau_d)e^{-\lambda_{\max}\tau_d}}{\lambda_{\min}^2}\:,
\]
then the attacker's best response to the action $\tau_d$ of the defender is $\lambda_a^*(\tau_d) = \lambda_{\min}$.
\end{theorem}

 \begin{theorem}\label{theorem6}
For the two person nonzero-sum game defined in Section \ref{sec:game} with the reward function in (\ref{eqn:linear_g_of_t}) and the exponentially distributed collocation time $\tau_a$ in (\ref{eqn:f_a}), if 
\[
C_d > \frac{1-(1+\lambda_{a}T)e^{-\lambda_{a}T}}{\lambda_a}\:,
\] 
then the defender's best response to the action $\lambda_a$ of the attacker is to stop migrations, i.e, $\tau_d^*(\lambda_a) = T$.

\end{theorem}

\section{Extended Game Model}\label{sec:extended}
In the aforementioned model, the attacker's goal is to be collocated with her victim as soon as possible before the victim is migrated. Evidently, upon collocation with her victim, the attacker will choose to reside there until $\tau_d$ since no detection mechanism is in place to urge her to evade. In this section, we extend the existing system model and consider the case in which the cloud data center is equipped with an intrusion detection system (IDS). The IDS monitors suspicious activities and captures malicious behavior of any user after a sufficient period of time $\delta_o$. For useful detection, $\delta_o < \tau_d$. Hence, the attacker may need to stop her collocation attacks before being detected. This introduces another control variable $s$ to be optimized by the attacker, namely how long she should continue to carry on the attack after successful collocation. The attacker does not know the operating threshold $\delta_o$ for the IDS -- otherwise, she would choose to stop right before the IDS threshold. However, the attacker is assumed to have prior knowledge of the distribution of $\delta_o$, in that he knows the pdf $y(\delta),\:\delta \in \left[0,T \right ]$. 

\begin{figure}
\centering
\includegraphics[width=0.95\linewidth,height=0.15\textheight]{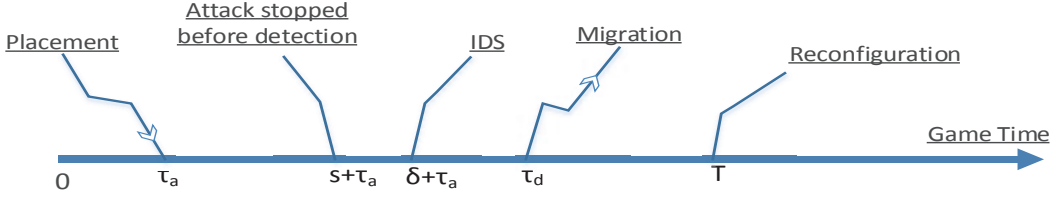}
\centering\caption{Attacker evades IDS by early stopping of malicious activity. 
}
\label{fig:system2}
\vspace{-0.6 cm}
\end{figure}

\begin{figure}
\centering
\includegraphics[width=0.95\linewidth,height=0.15\textheight]{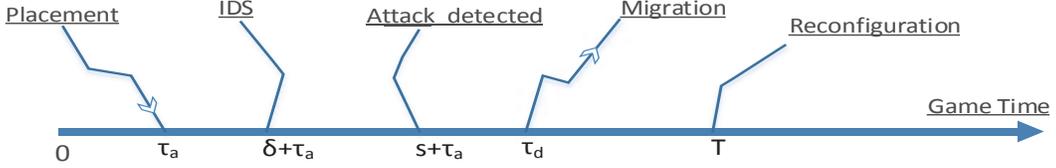}
\centering\caption{Attacker detected by the IDS.
}
\label{fig:system3}
\vspace{-0.6 cm}
\end{figure}

Next, we modify the attacker's payoff function $u_a$ in order to account for the probability of detection. In the event of detection, the attacker incurs a cost $D$ (since this user will be black-listed), but her gain is in the data leaked until detection. 
Therefore, we redefine the attacker's expected reward by averaging over both the detection threshold $\delta$ and the collocation time $\tau_a$ as,
\begin{equation}\label{eqn:ext_attacker}
\begin{split}
 u_a(\tau_d,\lambda_a,s) & = \frac{1}{\tau_d}    \int_{\delta = s}^{T} G(s) y(\delta) d \delta \int_{\tau_a = 0 }^{\tau_d - s} f_a (\tau_a;\lambda_a) d \tau_a \\ & + \frac{1}{\tau_d}    \int_{\delta = 0}^s (G(\delta) -D)\left (\int_{0}^{\delta_d-\delta}f_a (\tau_a;\lambda_a) d \tau_a\right) y(\delta) d \delta  \\ & + \frac{1}{\tau_d}  \int_{\delta = 0}^s \left( \int_{ \tau_d -\delta }^{\tau_d} G(\tau_d - \tau_a)f_a (\tau_a;\lambda_a) d \tau_a \right)y(\delta) d \delta \\ & + \frac{1}{\tau_d}  \int_{\delta = s}^{T} \left( \int_{ \tau_d -s }^{\tau_d} G(\tau_d - \tau_a)f_a (\tau_a;\lambda_a) d \tau_a \right)y(\delta) d \delta - \frac{ C_a \lambda_a }{\tau_d}   .
\end{split}
\end{equation}
\noindent
The first term in (\ref{eqn:ext_attacker}) accounts for the attacker's expected payoff in the event of no detection as the attacker stopped malicious activities before the IDS alarm, i.e, $ s < \delta $, as illustrated in Fig. \ref{fig:system2}. The second term represents the event of detection,  hence collocation ends at $\tau_a + \delta$, i.e., after a collocation duration $\delta$ as $ \delta < s$, as shown in Fig. \ref{fig:system3}. Therefore, the attacker incurs a detection loss $D$. The third and fourth terms account for the event of no detection but due to the migration mechanism. In other words, the attacker is not identified because $\tau_d-\tau_a < \min(\delta, s)$. The last term accounts for the cost of launching the attack. 

Similarly, we redefine the defender's expected payoff function,
\begin{equation}\label{eqn:ext_defender}
\begin{split}
 u_d(\tau_d,\lambda_a,s) & = \frac{-1}{\tau_d}    \int_{\delta = s}^{T} G(s) y(\delta) d \delta \int_{\tau_a = 0 }^{\tau_d - s} f_a (\tau_a;\lambda_a) d \tau_a \\ & - \frac{1}{\tau_d}    \int_{\delta = 0}^s (G(\delta) -D)\left (\int_{0}^{\delta_d-\delta}f_a (\tau_a;\lambda_a) d \tau_a\right)  y(\delta) d \delta  \\ & - \frac{1}{\tau_d}  \int_{\delta = 0}^s \left( \int_{ \tau_d -\delta }^{\tau_d} G(\tau_d - \tau_a)f_a (\tau_a;\lambda_a) d \tau_a \right)y(\delta) d \delta \\ & - \frac{1}{\tau_d}  \int_{\delta = s}^{T} \left( \int_{ \tau_d -s }^{\tau_d} G(\tau_d - \tau_a)f_a (\tau_a;\lambda_a) d \tau_a \right)y(\delta) d \delta - \frac{ C_d   }{\tau_d}\:.
\end{split}
\end{equation}

\section{Numerical Analysis}\label{sec:results}

\label{sec:num_analysis}
In this section, we provide numerical analysis of the proposed game model. 
To characterize the payoff functions for both players, we need to specify $G(t)$ and $f_a(\tau_a;\lambda_a)$. For the linear reward function $G(t)$ and the exponential density function $f_a(\tau_a;\lambda_a)$ described in (\ref{eqn:linear_g_of_t}) and  (\ref{eqn:f_a}), the reward functions can be readily expressed as
\begin{equation}\label{eqn:u_a}
u_a(\tau_d,\lambda_a) = \frac{\lambda_a \tau_d + e^{-\lambda_a \tau_d}-C_a \lambda_a^2 -1 } {\lambda_a \tau_d}, 
\end{equation}
\begin{equation}\label{eqn:u_d}
u_d(\tau_d,\lambda_a) = \frac{  1- \lambda_a \tau_d- e^{-\lambda_a \tau_d}-C_a \lambda_a} {\lambda_a \tau_d},
\end{equation}
for $\tau_d\in\mathcal{A}_d, \lambda_a\in\mathcal{A}_a$. 
In the following analysis, we study the behavior of the payoff functions for  both players. We illustrate the reward of the defender as a function of the migration time $\tau_d$ for a range of attack rates $\lambda_a$. For the attacker, we plot her reward as a function of $\lambda_a$ for different $\tau_d$. Afterwards, we investigate the effect of the migration cost $C_d$ and the attack cost $C_a$ on the reward functions and the existence of Nash equilibria. We also examine the best response curves for both players. We also show the regions of strict concavity which suffice for the existence of a Nash equilibrium. Finally, we generalize our analysis to investigate different scaling regimes of the reward function. 
\begin{figure}
\centering
\begin{minipage}{.45\textwidth}
  \centering
  \includegraphics[width= 3 in]{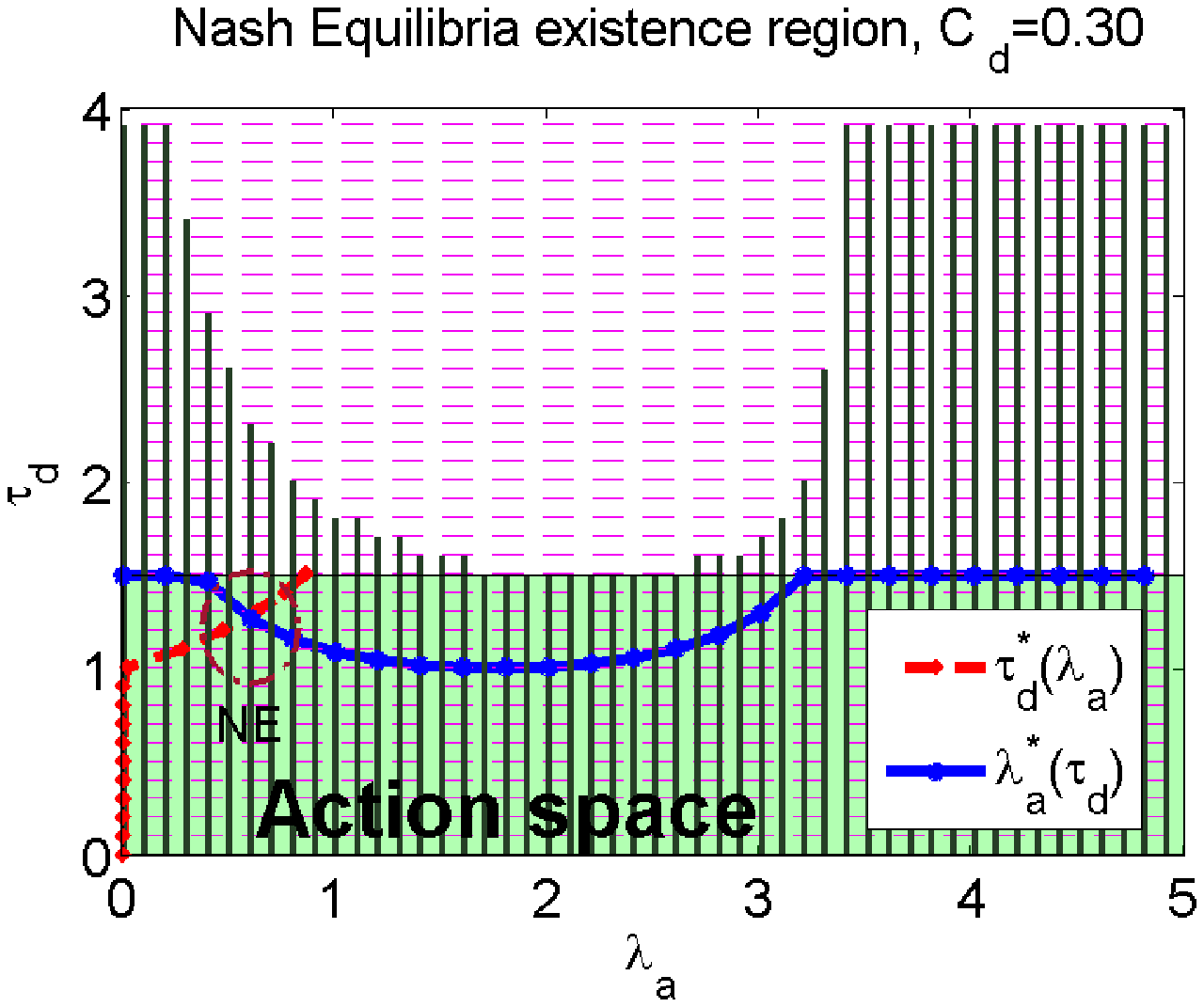}
 \centering\caption{NE existence region. Game admits a Nash equilibrium in pure strategies.} \label{fig:NE region}
\end{minipage}
\hspace{.05\linewidth}
\begin{minipage}{.45\textwidth}
  \centering
  \includegraphics[width = 3 in]{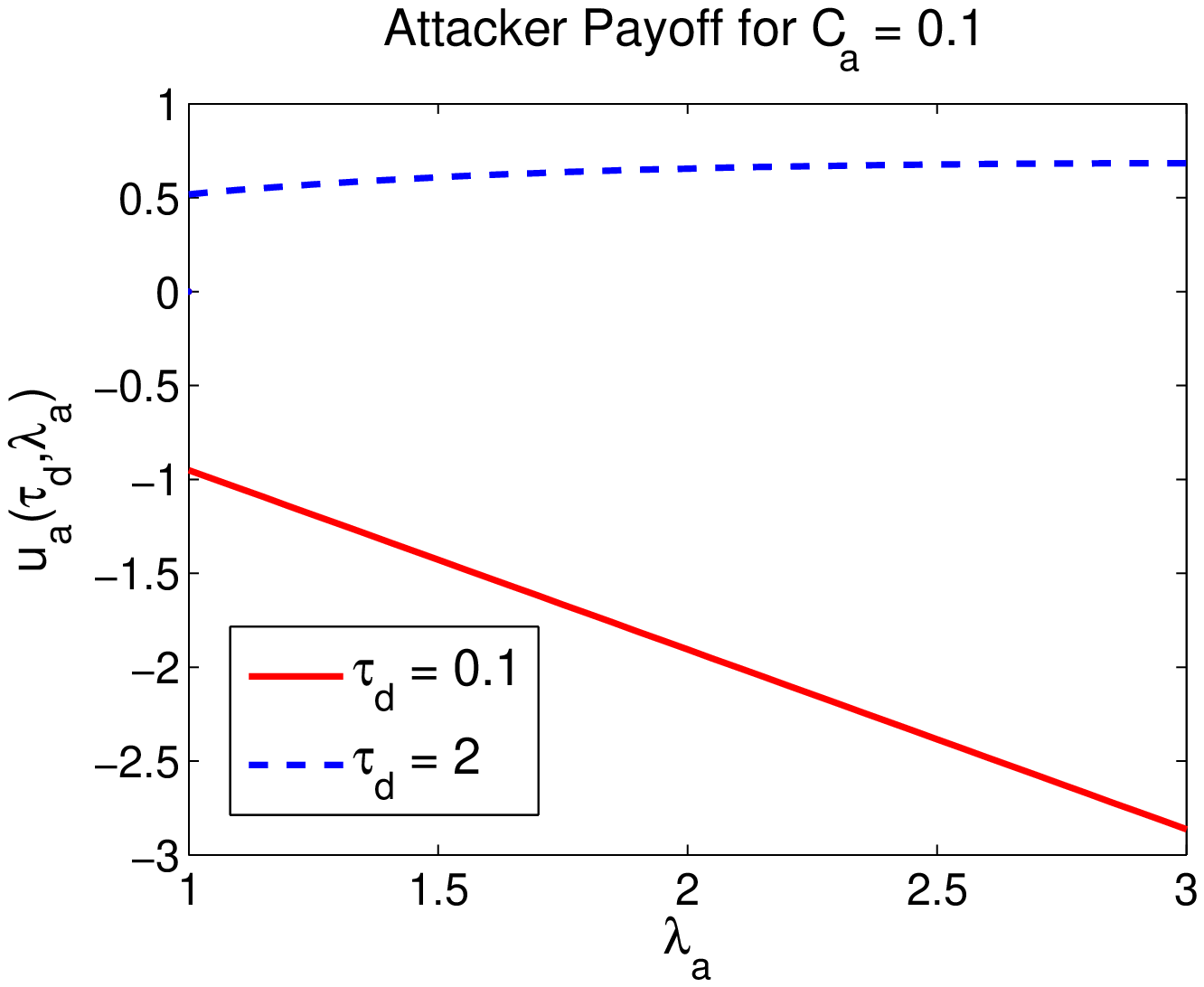}
 \centering\caption{At $\tau_d=0.1$, the attacker payoff is monotonically decreasing, but not for $\tau_d=2$, in agreement with the bound on $C_a$ in Theorem \ref{theorem5}.} \label{fig:AttacerMonotone}
\end{minipage}
\end{figure}   

We start our numerical analysis by reflecting on the theoretical analysis in Section \ref{subsec:special}. In Fig. \ref{fig:NE region}, we plot the NE existence region for $C_d = 0.3$. A NE exists if the conditions in the statement of Theorem \ref{theorem2} are satisfied to ensure strict concavity of both $u_a$ and $u_d$. Per Theorem \ref{theorem2}, at $T=1.5$ and $\lambda_{\max}=5$ as marked with the hashed rectangle, the game admits a Nash equilibrium in pure strategies. For the illustrated action space $\mathcal{A}$, the game satisfies the sufficient existence condition of Theorem \ref{theorem2}, i.e., the inequality $1-\lambda_a C_d < (1+\lambda_a \tau_d + \frac{\lambda^2_a \tau^2_d}{2})e^{-\lambda_a \tau_d}$ holds $\forall (\tau_d,\lambda_a) \in \mathcal{A}$. The figure illustrates the best response curves along with the game action space at $C_d = 0.3$ and $C_a = 0.5$. The horizontally dashed region is the region of concavity of $u_a$ in $\lambda_a$ for all $\tau_d\in\mathcal{A}_d$. Similarly, the vertically dashed region designates the region in which $u_d$ is concave in $\tau_d$ for every $\lambda_a\in\mathcal{A}_a$. As per Theorem \ref{theorem2}, any game defined within the region of intersection admits a NE in pure strategies. In this setting, the Nash equilibrium is unique -- shown as the unique intersection point of the best response curves for both players at $\tau^*_d = 1.27$ and $\lambda^*_a = 0.61$. Theorems \ref{theorem5} and \ref{theorem6} established lower bounds on $C_a$ and $C_d$ beyond which $u_a$ and $u_d$ are monotone. In Fig. \ref{fig:AttacerMonotone}, we numerically verify the monotonicity of $u_a$ for $\tau_d = 0.1$ and $\tau_d = 2$. Let $\lambda_{\min}=1$ and  $\lambda_{\max}=3$, hence according to Theorem \ref{theorem5}, $u_a$ is monotonically decreasing when $C_a > 0.04$ when $\tau_a = 0.1$. However, at $\tau_d =2$, the attack cost $C_a > 0.98$ ensures that $u_a$ is monotonically decreasing in $\lambda_a$. In Fig. \ref{fig:AttacerMonotone} where $C_a = 0.1$, it is shown that the corresponding $u_a$ is monotonically decreasing for all $\lambda_a \in \left[1,3 \right]$ for $\tau_d = 0.1$. At $\tau_d = 2$ when the condition on $C_a$ is not satisfied, the payoff $u_a$ is not monotonically decreasing. Next, we study and discuss the effect of different system parameters on the players' payoff and best response in comparison to other defense and attack policies.

\subsection{Payoff functions}
\begin{figure}
  \begin{subfigure}{.5\textwidth}
    \centering\includegraphics[width=7.5cm]{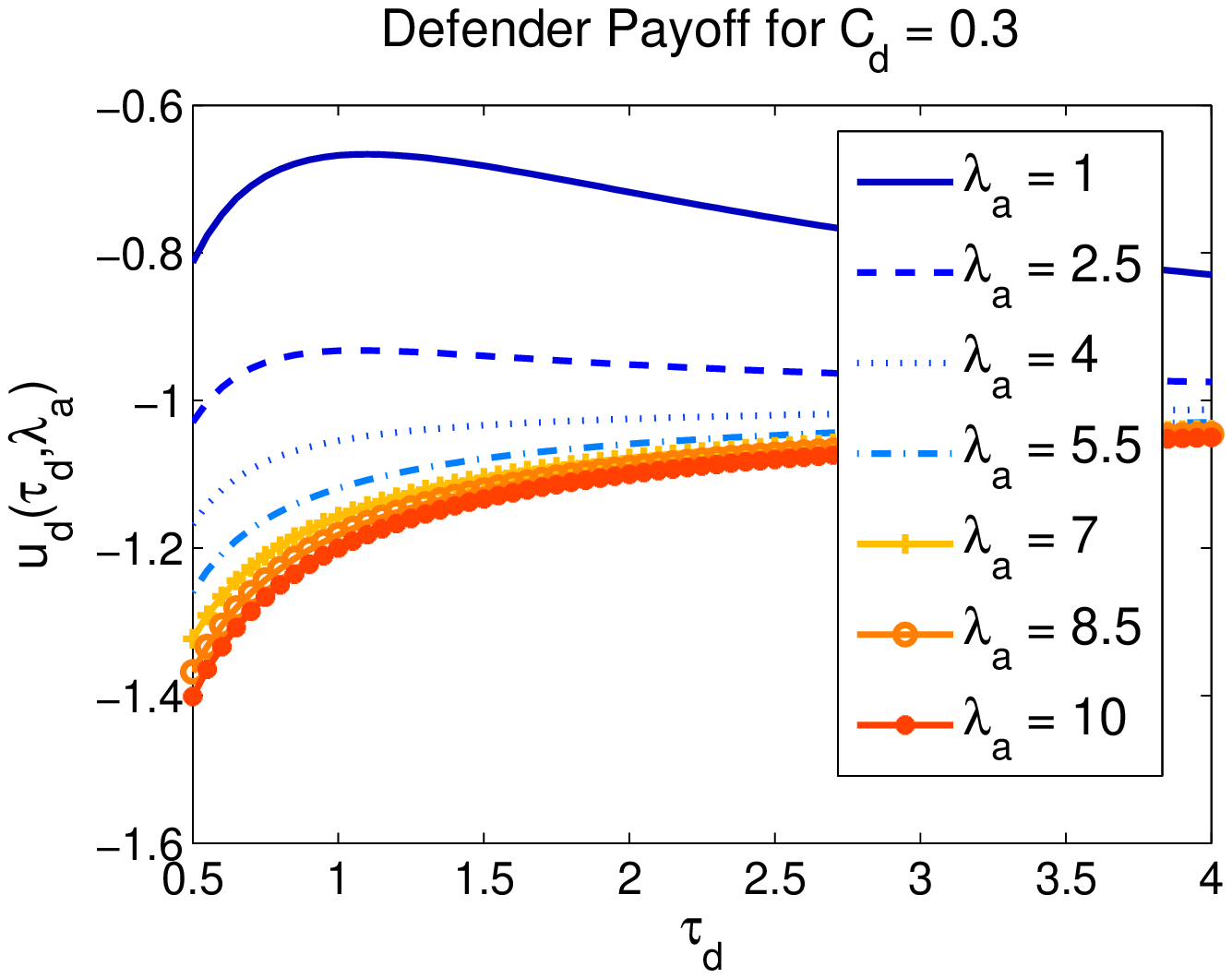}\caption{}\label{fig:defender reward2}
  \end{subfigure}
  \begin{subfigure}{.5\textwidth}
    \centering\includegraphics[width=7.5cm]{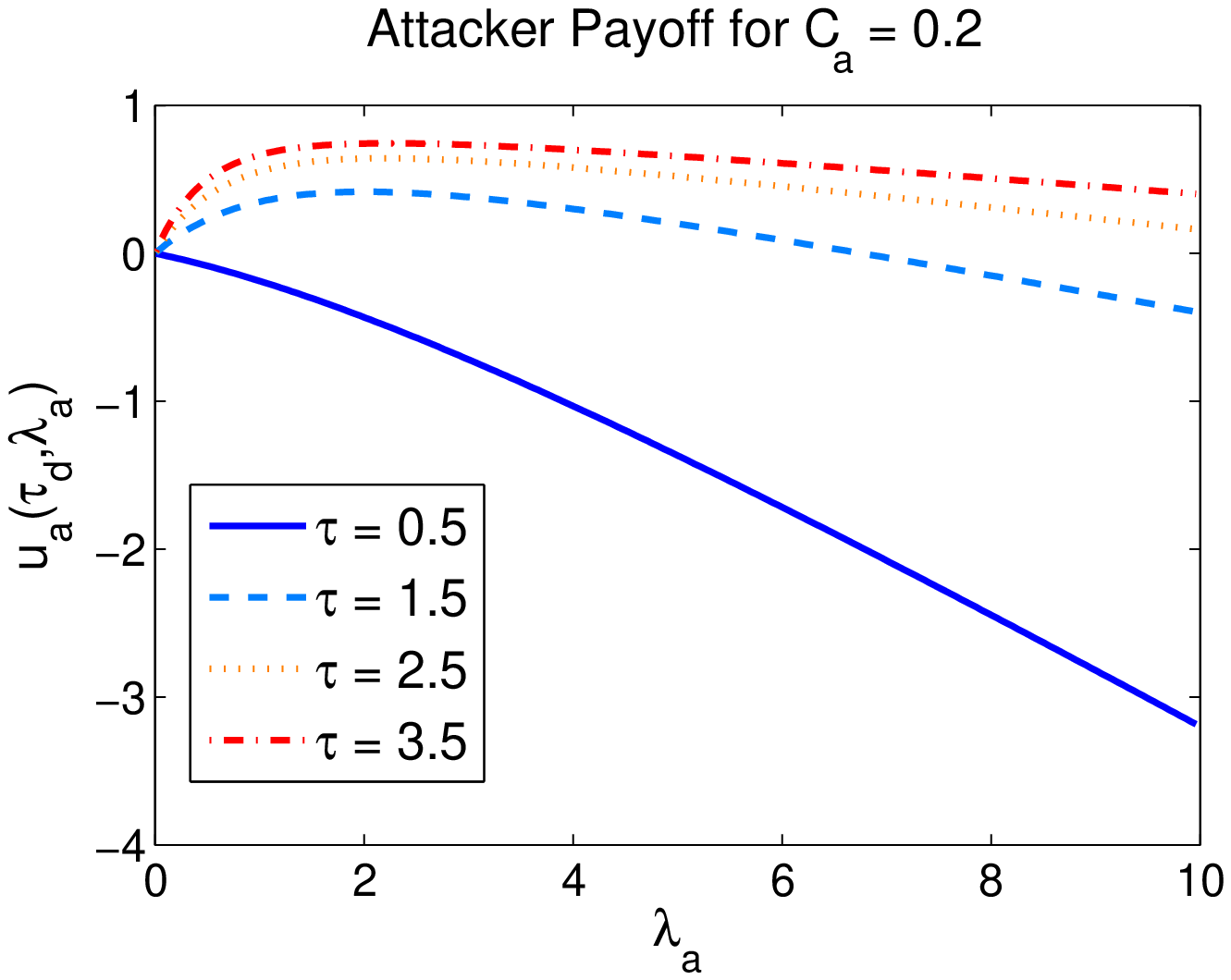}
    \caption{}\label{fig:attacker reward5}
  \end{subfigure}
  \caption{(a) Defender's reward versus migration time $\tau_d$; (b) Attacker's reward versus attack rate $\lambda_a$.}
\end{figure}

Fig. \ref{fig:defender reward2} shows the payoff function of the defender $u_d$ versus the migration time $\tau_d$ for $C_d = 0.3$ and $T=4$. The figure highlights the tradeoff faced by the defender as he seeks to optimize $\tau_d$ to secure the system through VM migration while avoiding a large migration overhead. Evidently, the optimal migration time $\tau_d^*$ depends on the attacker's strategy $\lambda_a$. The tradeoff shown in Fig. \ref{fig:defender reward2} agrees with our intuition based on the game model. Specifically, a very small $\tau_d$ signifying a high VM migration rate is associated with a high migration cost that dominates the payoff function $u_d$. On the other hand, with a larger $\tau_d$, the VMs dwell for a longer period of time on the same physical node giving the attacker more room to collocate and leak data from his targeted VM. In Fig. \ref{fig:defender reward2}, we compare the defender's reward at different attack rates $\lambda_a$. It is clear that when the attack is less aggressive, the defender is able to maximize his payoff by reducing the migration time $\tau_d$ at the expense of higher migration cost. Therefore, when $\lambda_a$ increases from $1$ to $2.5$, the optimal $\tau_d$ reduces from $1.25$ to $0.85$. However, when the attacker is very aggressive (using a high attack rate), the defender is better off avoiding the migration cost by increasing $\tau_d$ to its maximum permissible value, i.e., $\tau_d^* = T$. This means that no migration rate would limit the damage of the attacker, thus the best course of action in this case is to migrate at the end of cycle $T$ to avoid the migration cost.

In Fig. \ref{fig:attacker reward5}, we plot the attacker's expected payoff $u_a$ versus the attack rate $\lambda_a$ for different defense actions $\tau_d$ for an attack cost $C_a = 0.2 $. As shown, the optimal attack rate depends on the defender's action. As the attack rate increases, the cost of attack increases and becomes the dominant term in the payoff function. Moreover, as the defender reduces his time to migrate $\tau_d$, the attacker's reward decreases. This is due to the fact that when $\tau_d$ is small (a higher migration rate), there is a shorter time window for the attacker to successfully collocate with her victim. Contrariwise, when the migration rate is not too high (i.e., $\tau_d$ is fairly large), the attacker can maximize his reward by increasing the attack rate $\lambda_a$. This can be seen in Fig. \ref{fig:attacker reward5} where $\tau_d$ is reduced from $\tau_d = 3.5$ to $\tau_d = 1.5$, and the optimal attack rate that maximizes the payoff $u_a$ increases from $\lambda_a^* = 1.7$ to $ \lambda_a^* = 2$. However, if the defender is migrating the VMs at a very high rate, i.e, $\tau_d$ is very small, the attacker's best response is to attack at the minimum possible rate or completely back-off since the attack is useless. To better understand the effect of the migration (attack) cost on the optimal migration (attack) rate for the defender (attacker), in the following two subsections we study the behavior of the payoff functions at different values of the cost. We also study the behavior of the best response curves to gain more insight into the tradeoffs associated with this game.    
\subsection{Cost effect and monotonicity}\label{subsec:mono}

\begin{figure}
  \begin{subfigure}{.5\textwidth}
    \centering\includegraphics[width=7.5cm]{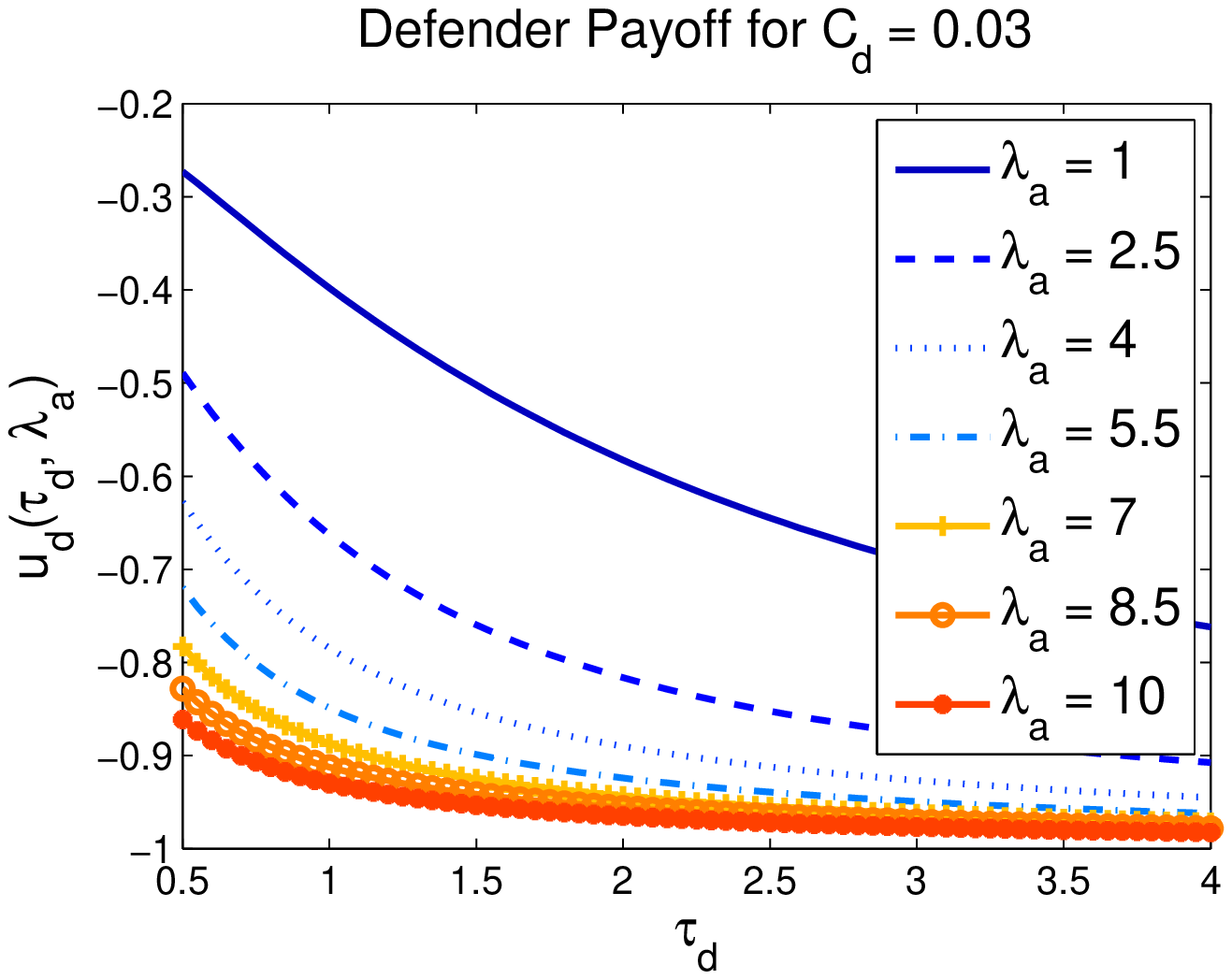}\caption{}\label{fig:defender reward1}
  \end{subfigure}
  \begin{subfigure}{.5\textwidth}
    \centering\includegraphics[width=7.5cm]{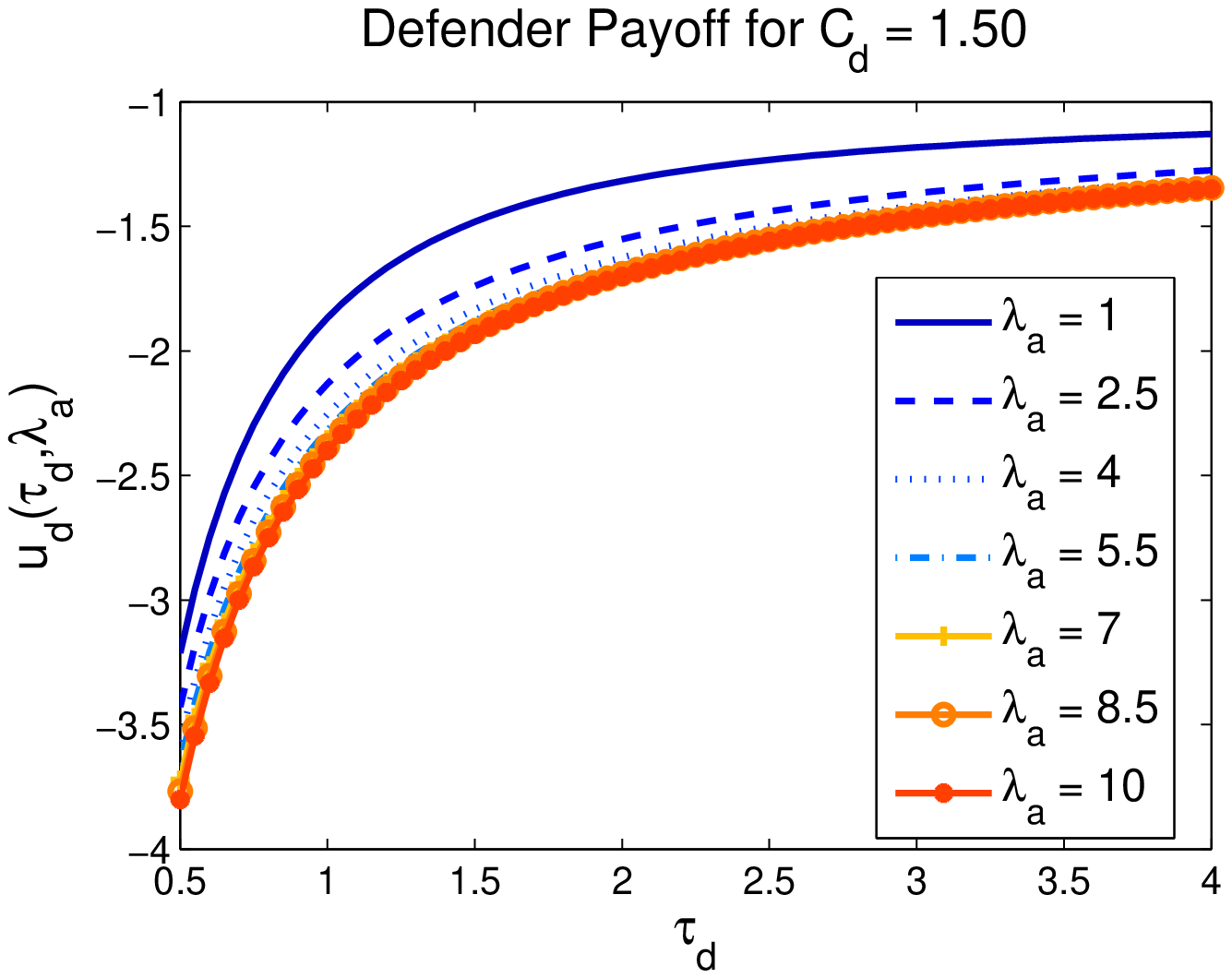}
    \caption{}\label{fig:defender reward4}
  \end{subfigure}
  \caption{Defender's reward versus migration time $\tau_d$ for (a) $C_d = 0.03$, and (b) $C_d = 1.5$.}
\end{figure}

To show the effect of the migration and attack costs $C_d$ and $C_a$, we plot the players reward functions for different values of the cost. In Fig. \ref{fig:defender reward1}, we plot the defender's payoff versus $\tau_d$ for different attack strategies for a a fairly small migration cost is $C_d = 0.03$. At this small migration cost, the defender's best response is to always migrate at the highest permissible rate, i.e, $\tau_d^* = \tau_{\min}$ regardless of the attack rate $\lambda_a$. Hence, the leakage loss term dominates the defender's payoff function $u_d$ at this small migration cost. Indeed, referring to (\ref{eqn:u_d}), $u_d$ is monotonically decreasing in $\tau_d$ when $C_d \to 0$. On the other hand, when the migration cost is high as shown in Fig. \ref{fig:defender reward4} where $C_d = 1.5$, the defender's best response is $\tau_d^* = T$ to reduce the associated migration cost. We remark that the reward function is monotonically increasing in $\tau_d$ for such high migration cost, a fact which was established analytically in Theorem \ref{theorem6}.  

\begin{figure}
  \begin{subfigure}{.5\textwidth}
    \centering\includegraphics[width=7.5cm]{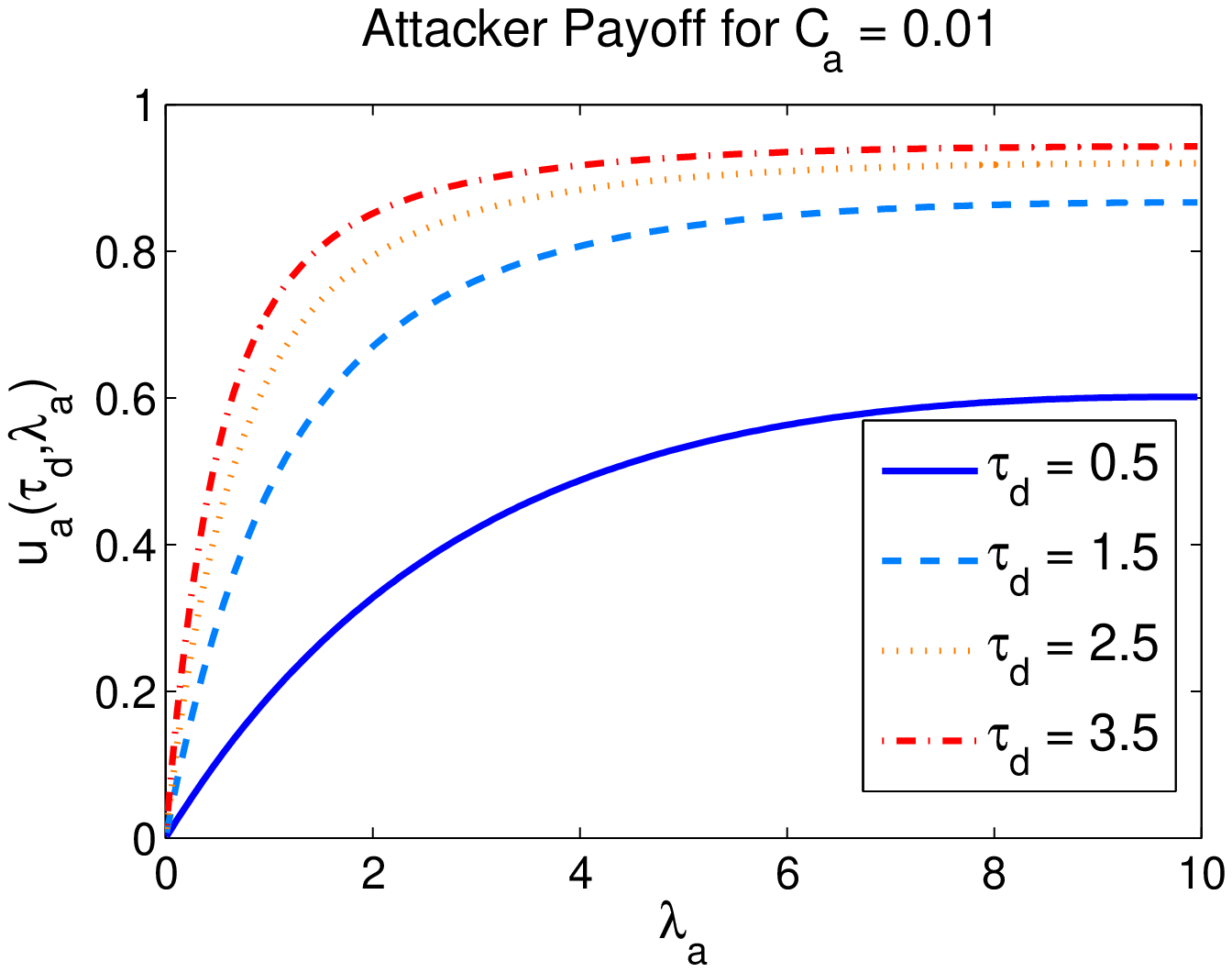}\caption{}\label{fig:attacker reward1}
  \end{subfigure}
  \begin{subfigure}{.5\textwidth}
    \centering\includegraphics[width=7.5cm]{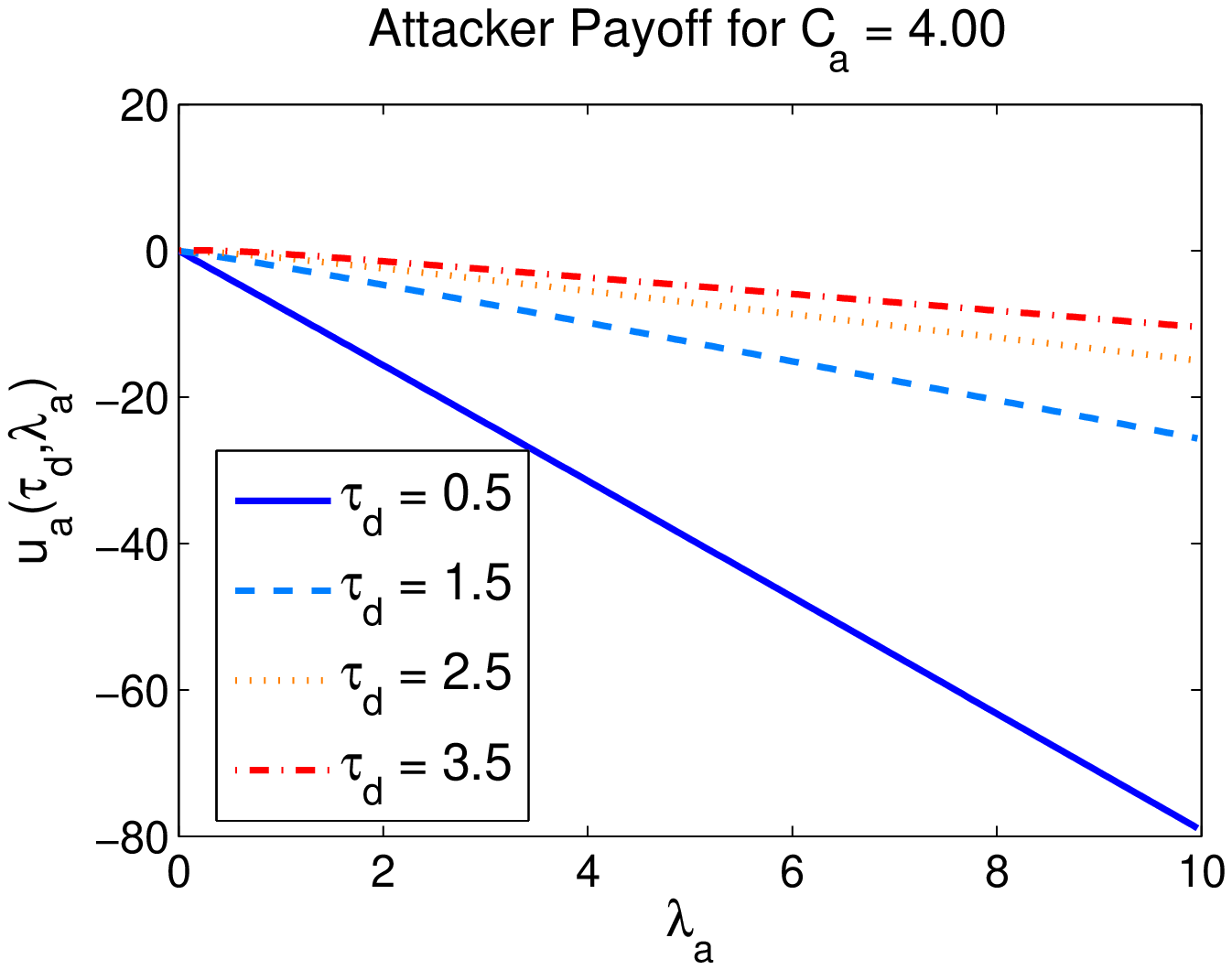}
    \caption{}\label{fig:attacker reward3}
  \end{subfigure}
  \caption{Attacker's reward versus attack rate $\lambda_a$ for (a) $C_a = 0.01$, and (b) $C_a = 4$.}
\end{figure}

Similarly, the effect of the attack cost $C_a$ can be shown in Fig. \ref{fig:attacker reward1}. At a very small attack cost, $C_a = 0.01$, as shown in Fig. \ref{fig:attacker reward3}, the attacker's best attack strategy is to attack aggressively at the $\lambda_{\max}$ to maximize the chances of successful collocation regardless of the defender's action. Recalling the attacker's payoff function in (\ref{eqn:u_a}), $u_a$ is monotonically increasing in $\lambda_a$ when $C_a \to  0$. In case of a high attack cost, the behavior of the payoff function is reversed as shown in Fig. \ref{fig:attacker reward3} where $C_a = 4$. In this case, the cost of the attack term dominates the payoff function. Therefore, the best action for the attacker is $\lambda_{\min}$ regardless of the action of the defender. This behavior is confirmed by the analysis in Theorem \ref{theorem5}. 

\subsection{Best response curves}\label{subsec:BR}

\begin{figure}
  \begin{subfigure}{.5\textwidth}
    \centering\includegraphics[width=7.5cm]{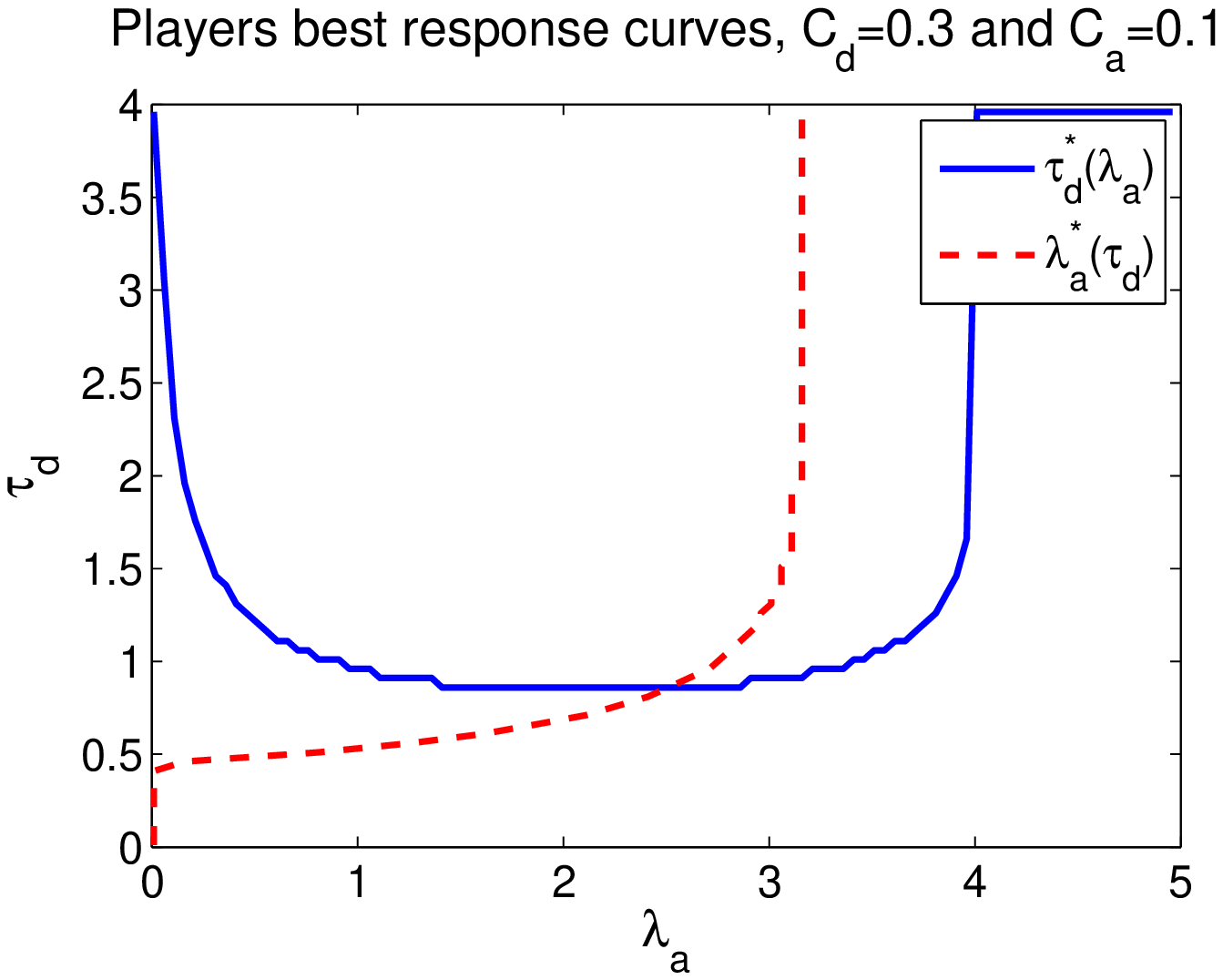}\caption{}\label{fig:BR_saddle2}
  \end{subfigure}
  \begin{subfigure}{.5\textwidth}
    \centering\includegraphics[width=7.5cm]{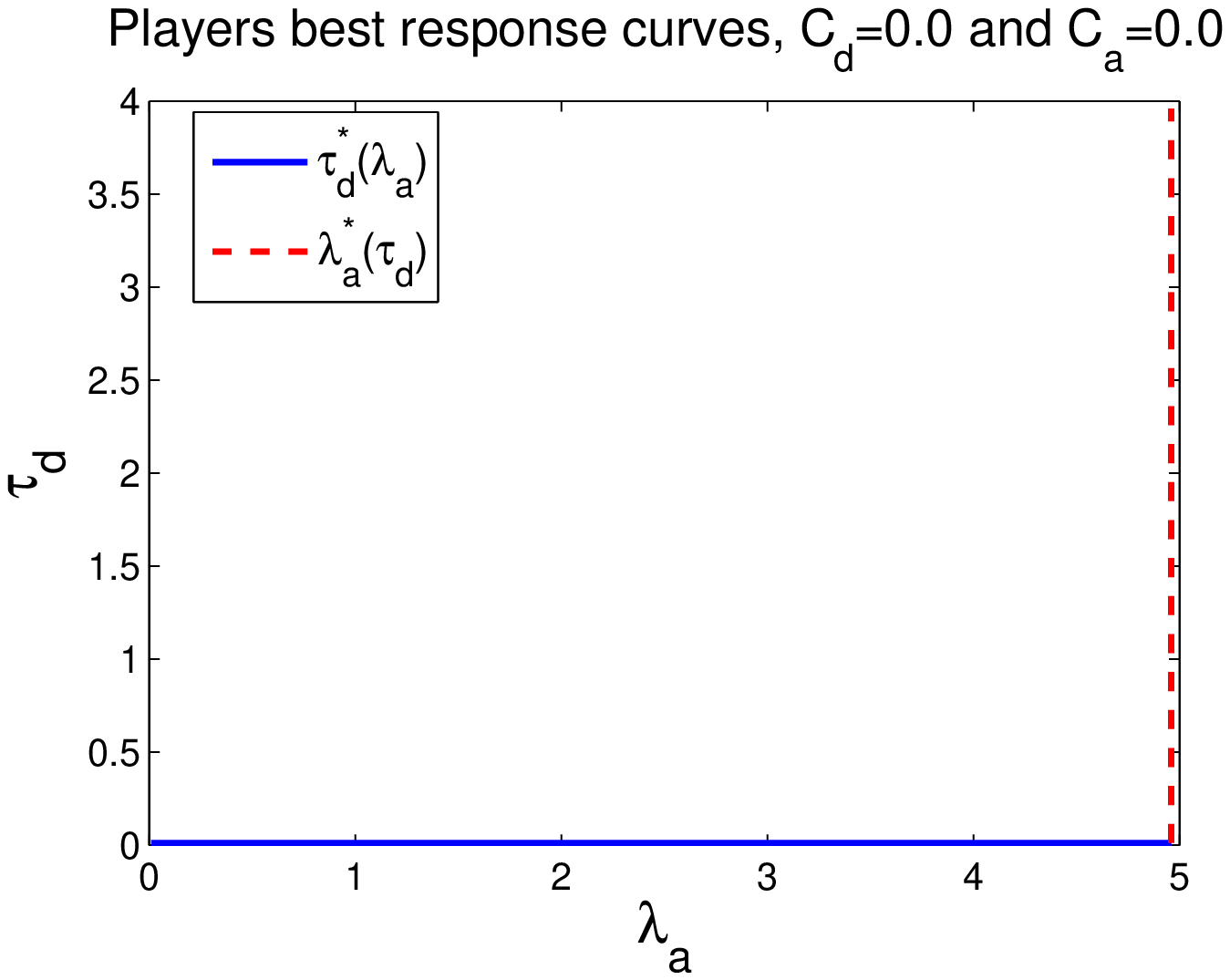}
    \caption{}\label{fig:BR_no_cost}
  \end{subfigure}
  \begin{subfigure}{.5\textwidth}
    \centering\includegraphics[width=7.5cm]{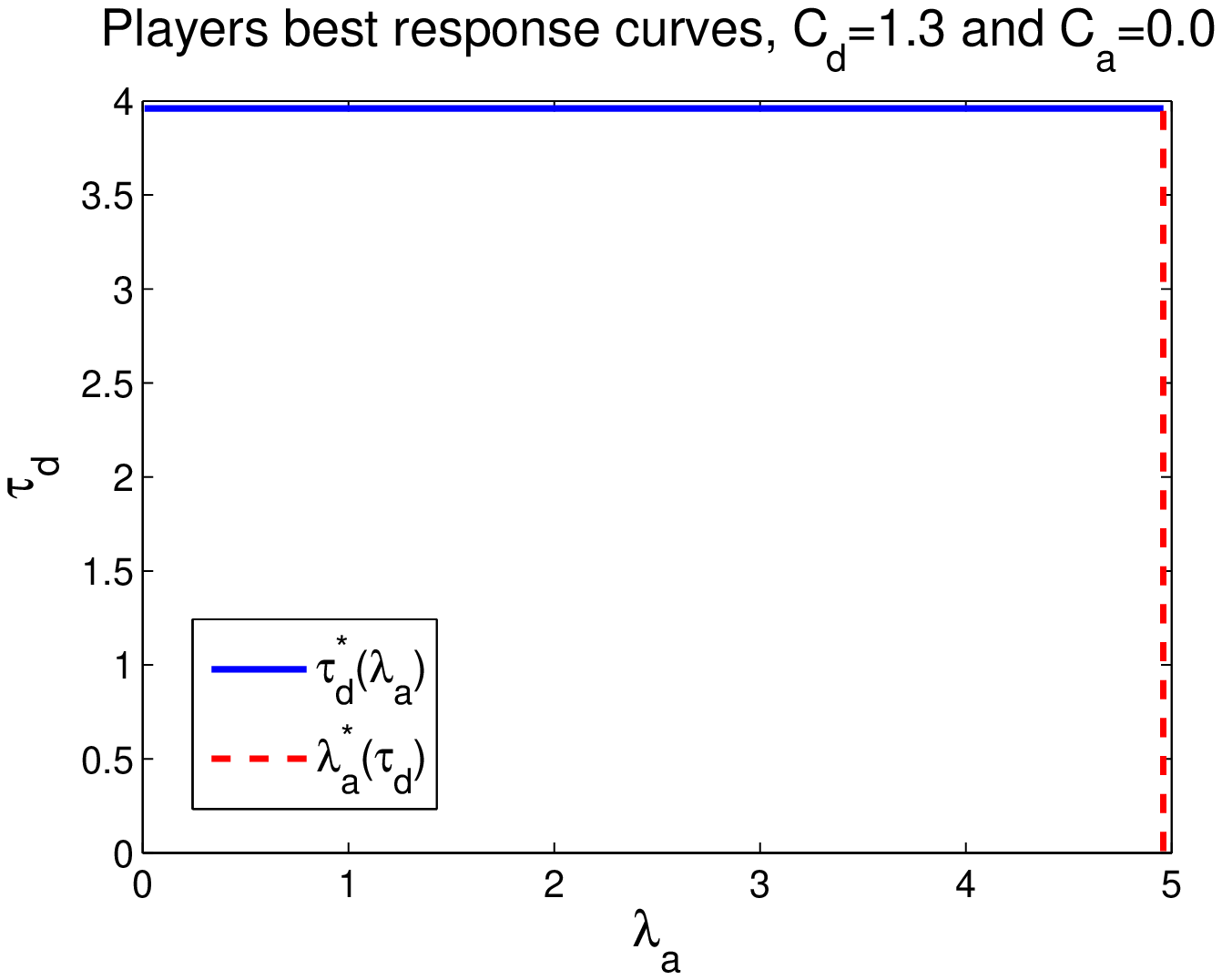}
    \caption{}\label{fig:BR_defender_tradeoff2}
  \end{subfigure}
  \begin{subfigure}{.5\textwidth}
    \centering\includegraphics[width=7.5cm]{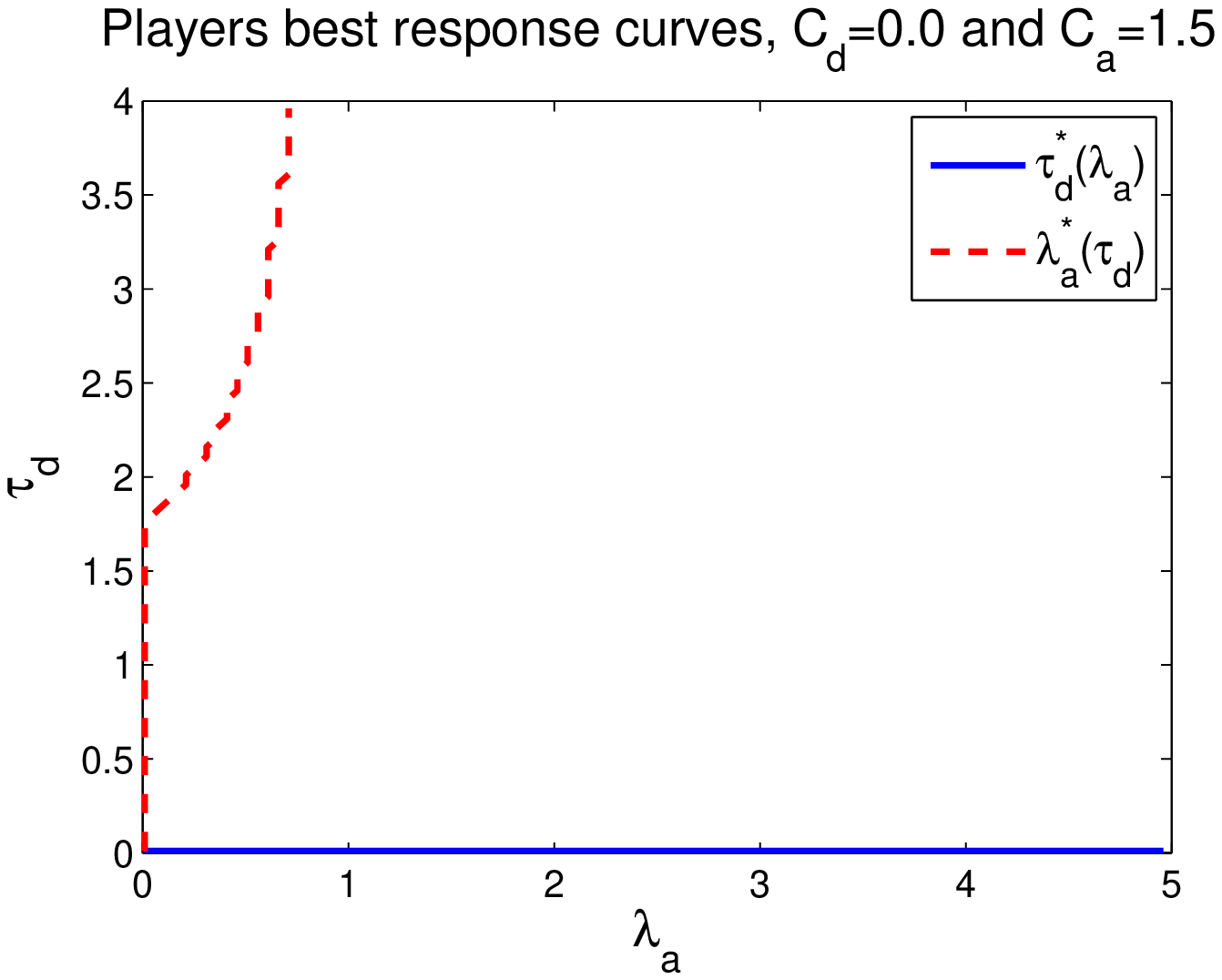}
    \caption{}\label{fig:BR_attacker_tradeoff}
  \end{subfigure}
  \caption{Players best response curves for different cost values.}
\end{figure}

In this section, we study the best response curves for both players based on Definition \ref{def:br} to provide more insight into the optimal action of a player as function of the action of the opponent. The solid blue line in Fig. \ref{fig:BR_saddle2} shows the defender's best response curve $\tau^*_d$ as function of $\lambda_a$. The attacker's best response curve $\lambda^*_a$ as function of the defender's action $\tau_d$ is shown in dashed red line. In this scenario, we set $T = 4$, $\lambda_{\max} = 5$, $C_d = 0.3$, and $ C_a = 0.1$. In Fig. \ref{fig:BR_saddle2}, the intersection point of the two response curves is the unique Nash equilibrium. The point(s) of equilibria depend on the values of $C_a$ and $C_d$ as detailed next. The best response curves also underscore the tradeoff for each player. For example, at equilibrium the defender migrates with $\tau_d = 0.86 $ while the attacker uses rate $\lambda_a = 2.51$ for the attack. Clearly, at low attack rate, VM migration at a very small migration rate, i.e, larger $\tau_d$, is more favorable. As the attack rate increases, the defender is urged to migrate the VMs at faster rate, wherefore $\tau^*_d$ decreases as $\lambda_a$ increases, but only until a certain point where faster migration becomes futile. Indeed, when the attack rate is overwhelming, it is more rewarding for the defender to use a large $\tau_d$ to alleviate high migration costs. 
On the attacker's side, a similar tradeoff is observed. The attacker attacks the system at the minimum rate $\lambda_{\min}$ as long as the VM stays on the same physical node for a duration $\tau_d < $ 0.4 since it is very hard to collocate when migration is taking place at such high rates. If the defender increases the time before migrating, i.e $\tau_d > 0.4$, the attacker is enticed to attack the system at higher rates to increase leakage. However, the maximum attack rate the attacker will select is $\lambda_a = 3.16$ which is strictly smaller than $\lambda_{\max}$ since the resultant attack cost yields a smaller overall payoff. 
The best response curves also demonstrate the monotonicity of the payoff functions with respect to $C_a$ and $C_d$ as explained earlier in Section \ref{subsec:mono}. To show this, Fig. \ref{fig:BR_no_cost}, \ref{fig:BR_defender_tradeoff2} and \ref{fig:BR_attacker_tradeoff} illustrate the best response curves at extreme cost values. In particular, in Fig. \ref{fig:BR_no_cost}, both $C_d$ and $C_a$ are set to zero. It is obvious that the defender is migrating with the highest permissible frequency such that, $\tau^*_d = \tau_{\min}$ for any attack rate. In response, the attacker's best action is $\lambda_a = \lambda_{\max}$ regardless of the defender's action. Hence, when the costs of migration and attack are zero, both players do not face any tradeoffs and the game is zero-sum. 
Fig. \ref{fig:BR_defender_tradeoff2} shows another extreme scenario where only the defender faces a very high cost for migration. His best response is $\tau^*_d = T= 4$, which corresponds to the lowest migration rate possible. 
In Fig. \ref{fig:BR_attacker_tradeoff}, the attack cost $C_a =1.5$ while the defender incurs zero cost for migration. Hence, the defender adopts the highest migration rate at $\tau_d^* =\tau_{\min}$ against any attack rate. In response, it is more rewarding for the attacker to attack at $\lambda_{\min}$ unless the defender does not migrate before $\tau_d = 1.8$, in which case the attacker would increase the attack rate, i.e., $\lambda_a > \lambda_{\min}$.  
 
\begin{figure}
  \begin{subfigure}{.5\textwidth}
    \centering\includegraphics[width=7.5cm]{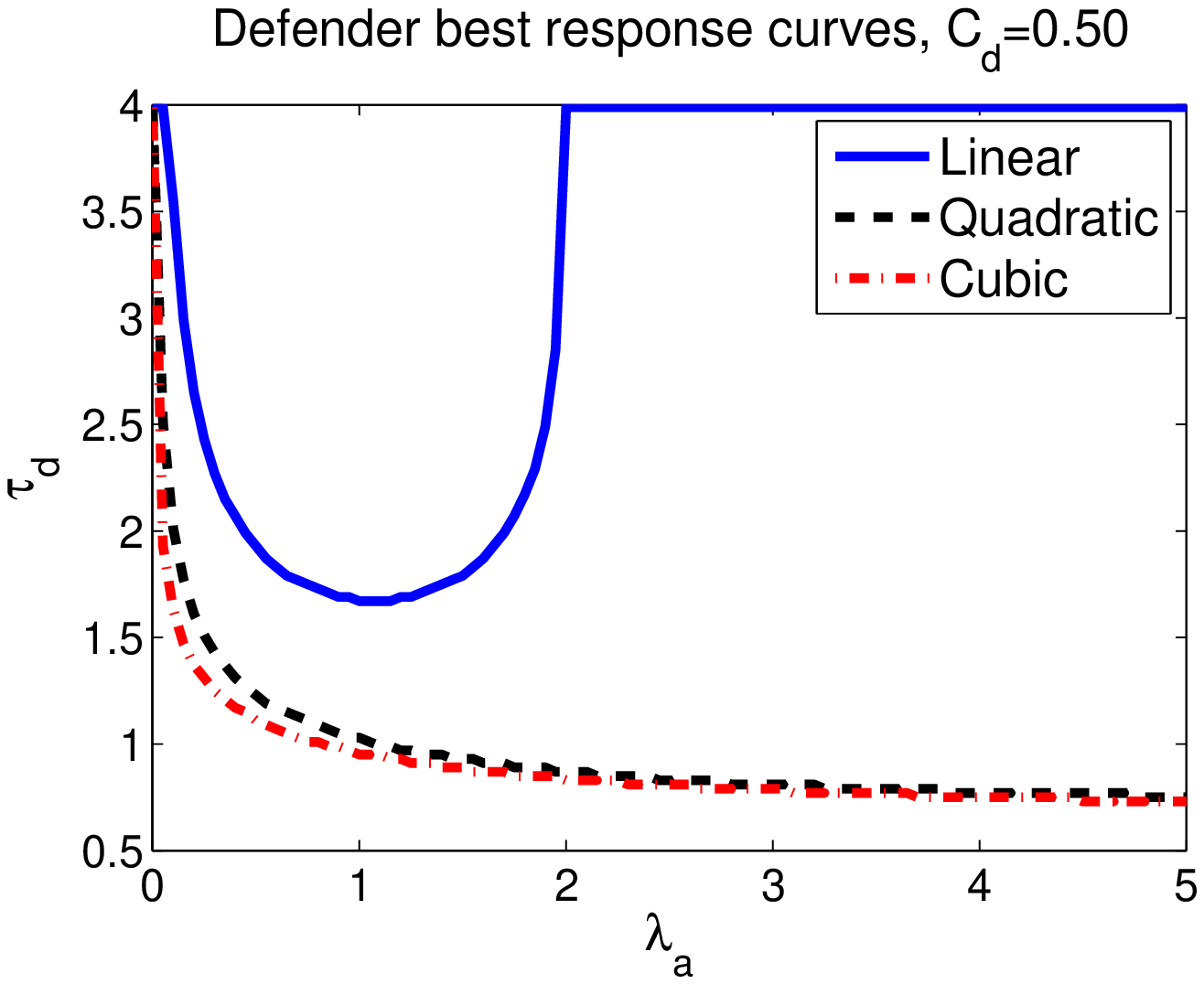}\caption{}\label{fig:extreme_regimes_defender}
  \end{subfigure}
  \begin{subfigure}{.5\textwidth}
    \centering\includegraphics[width=7.5cm]{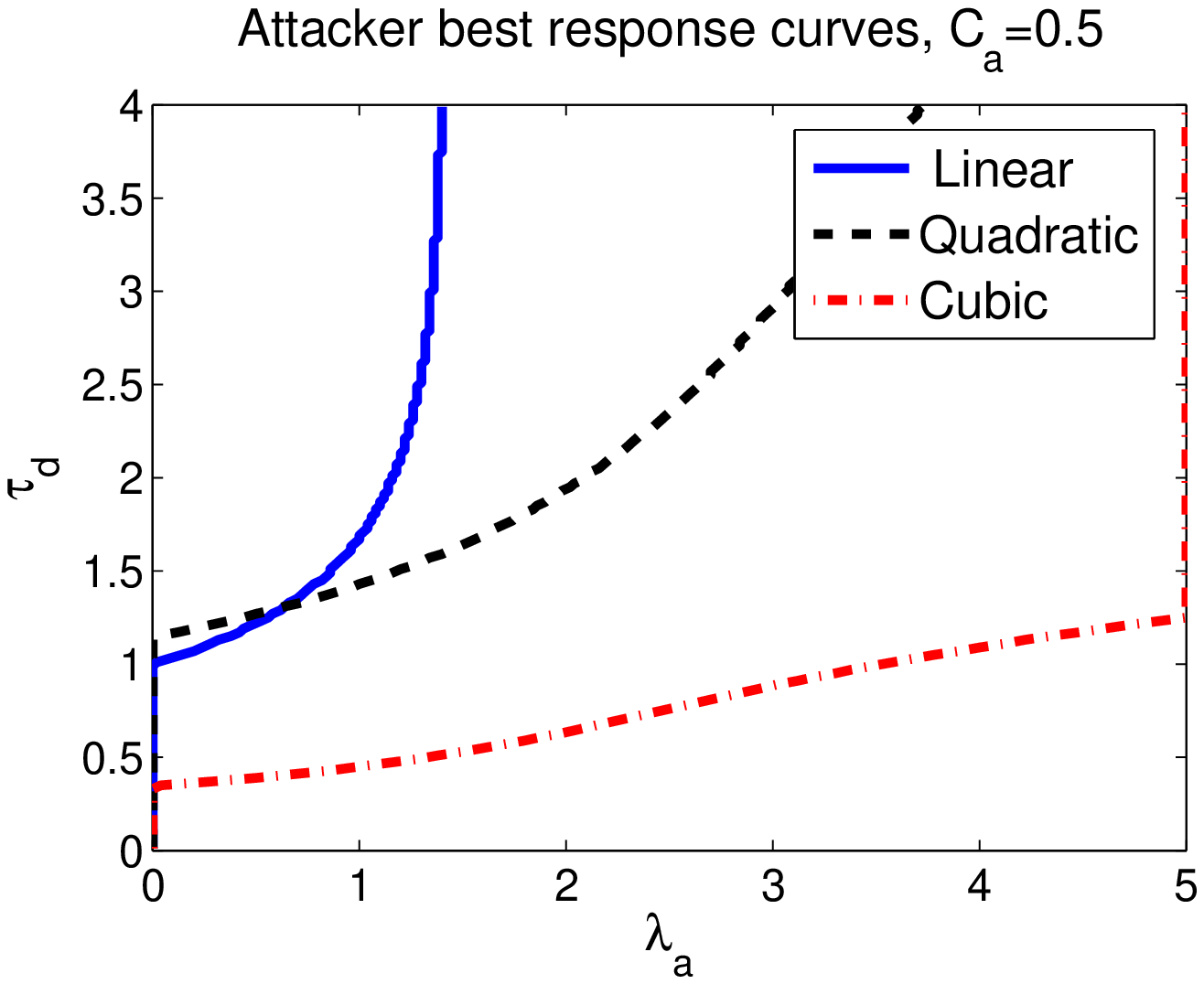}
    \caption{}\label{fig:extreme_regimes_attacker}
  \end{subfigure}
  \caption{Defender's (a) and Attacker's (b) best response curves for different reward scaling regimes.}
\end{figure}

\subsection{Different reward scaling regimes}
\label{subsec:G_scalings}

In the numerical analysis above, we considered the reward function $G$ to be linearly increasing in the collocation duration. In this section, we study other scaling regimes. In particular, we consider the scenario where the reward $G(\tau_d, \tau_a)$ scales quadratically or cubically with the collocation duration. In Fig. \ref{fig:extreme_regimes_defender}, we plot the defender's best response curves for 
linear, quadratic, and cubic reward functions. Intuitively, higher order reward functions are more disposed to dominate the payoff functions than for the linear scaling.  
In Fig. \ref{fig:extreme_regimes_defender}, the migration cost is set to $C_d = 0.5$. For the linear regimes, the defender is facing exactly the same tradeoff discussed earlier in Section \ref{subsec:BR}. However, for higher order reward regimes, the reward term dominates the payoff over the entire range of attack rates in this case. Therefore, the defender is consistently urged to increase the migration rate as the attacker increases her attack rates.
With the quadratic and cubic reward functions the defender's best response is shown to exhibit a similar behavior, but conceivably the cubic reward yields a higher increase in the rate of migration.

In Fig. \ref{fig:extreme_regimes_attacker}, the attacker's best response curves are plotted for different reward functions. The higher the order of the reward regime, the more is the attacker enticed to attack. In the linear regime, the attacker's best response rate is non-vanishing and increasing in $\tau_d$ for $\tau_d > 1$, but saturates at $\lambda_a = 1.4$ as soon as the cost of the attack starts to dominate the attacker's payoff. For both the quadratic and cubic regimes, the higher reward from data leakage entices the attacker to attack at higher rates as $\tau_d$ increases.   

As shown in Fig. \ref{fig:extreme_regimes_attacker}, the cubic regime is extremely rewarding to the attacker, and as a result the attacker affords to attack at the maximum permissible rate as the reward term dominates her payoff function. 
\subsection{Simulation of the game}
In this section, we compare the payoff of both players playing NE strategies to the payoffs of other defense and attack strategies. As per our theoretical analysis in Section \ref{sec:analysis}, the players' optimal (Nash equilibrium) policies depend on the values of the associated costs $C_d$ and $C_a$. Table \ref{table1} presents the results of a simulation of the game for the linear reward regime in which $G(\tau_d, \tau_a) = (\tau_d - \tau_a)^+ = \max(\tau_d - \tau_a,0)$ at different values of $C_d$ and $C_a$. For the numerical results, the maximum collocation time  is set to $T=3$ and the maximum attack rate is $\lambda_{\max} = 3$. The results underscore that a rational attacker would adapt the attack rate to the attack cost to avoid incurring high cost and/or launching useless attacks. For example, when $C_a = 6$, the payoff corresponding to the NE converges to that of the No Attack strategy, which is substantially higher than the payoff of an aggressive attacker of $-1.8\times 10^3$ due to a substantial attack cost.  
Similarly, the defender should not resort to very high frequency migration (equivalently, small $\tau_d$), unless the migration cost is fairly low. For example, the results in the table show that the payoff of the defender adopting the NE policy tends to that of the No defense policy as $C_d$ increases.
The last column designated as worst case (for the defender) corresponds to the scenario where the attacker is attacking at the highest rate while the defender does not adopt any migration policy. 
The loss of the defender for not migrating compared to he NE strategy is more pronounced when the NE point has $\tau_d^* < T$, i.e., when the defender is in a position to defend the system through VM migrations.  

 \begin{table}
 \scriptsize
\centering
\begin{tabular}{cc|cc|cc|cc|cc|cc}
\hline
\multicolumn{2}{c}{Cost} & \multicolumn{2}{c}{NE} 
&
\multicolumn{2}{c}{No Defense} &
\multicolumn{2}{c}{No Attack} &
\multicolumn{2}{c}{Aggressive Attack} &
\multicolumn{2}{c}{Worst case} \\\cline{1-12} 
$C_d$ & $C_a$ & $u_d$& $u_a$    & $u_d$ & $u_a$   & $u_d$ & $u_a$   & $u_d$ & $u_a$   & $u_d$ & $u_a$      \\
\hline
0	&0	&-1.49E-02	&1.49E-02	&-0.8893	&0.8893	&-5.00E-05	&5.00E-05	&-1.49E-02	&1.49E-02	&-0.8896	&0.8896 \\
0.1	&0	&-0.6672	&0.3894	&-0.9229	&0.8896	&-0.2796	&0.0018	&-0.6672	&0.3894	&-0.9229	&0.8896 \\
0.1	&0.1	&-0.4146	&0.0204	&-0.72	&0.6532	&-0.1986	&5.85E-04	&-0.685	&-0.1013	&-0.9229	&0.7896 \\
0	&0.1	&-5.00E-05	&-0.1	&-0.0149	&0.0146	&-5.00E-05	&-0.1	&-1.49E-02	&-30.0851	&-0.8896	&0.7896\\
0.4	&0.2	&-0.9513	&0.4201	&-0.968	&0.7015	&-0.2724	&0.0062	&-1.0472	&0.3836	&-1.0225	&0.6896 \\
0.4	&0.4	&-0.7504	&0.1593	&-0.8197	&0.5526	&-0.3009	&0.0038	&-1.0539	&-0.1255	&-1.0225	&0.4896 \\
0.4	&0.6	&-0.6319	&0.1047	&-0.6986	&0.4341	&-0.2724	&0.0035	&-1.0472	&-0.4137	&-1.0225	&0.2896 \\
0.8	&0.6	&-0.9984	&0.4914	&-0.9984	&0.4914	&-0.2807	&0.0129	&-1.1554	&0.2896	&-1.1554	&0.2896 \\
0.8	&1	&-0.8904	&0.2963	&-0.8914	&0.3566	&-0.3086	&0.0097	&-1.1726	&-0.2333	&-1.1554	&-0.1104 \\
2	&4	&-0.7496	&0.0054	&-0.7496	&0.0054	&-0.6794	&0.0016	&-1.5541	&-3.1104	&-1.5541	&-3.1104 \\
3	&0	&-1.8889	&0.8889	&-1.8859	&0.8893	&-1.0149	&0.0149	& -1.8893	& 0.8893	&-1.8863	&0.8896
 \\
 0	&6	&-4.9998e-05
	&-6.0	& -0.0149	&-0.0050	&-4.9998e-05	&-6.0000& -0.0149	& -1.8060e+03	&-0.8896 &-5.1104
 \\
 \cline{1-12}
\end{tabular}
\caption{Players' payoff for several attack and defense strategies.}\label{table1}
\end{table}

\subsection{Extended model with IDS}
\begin{figure}
\centering
\begin{minipage}{.45\textwidth}
  \centering
\includegraphics[width=2.5 in]{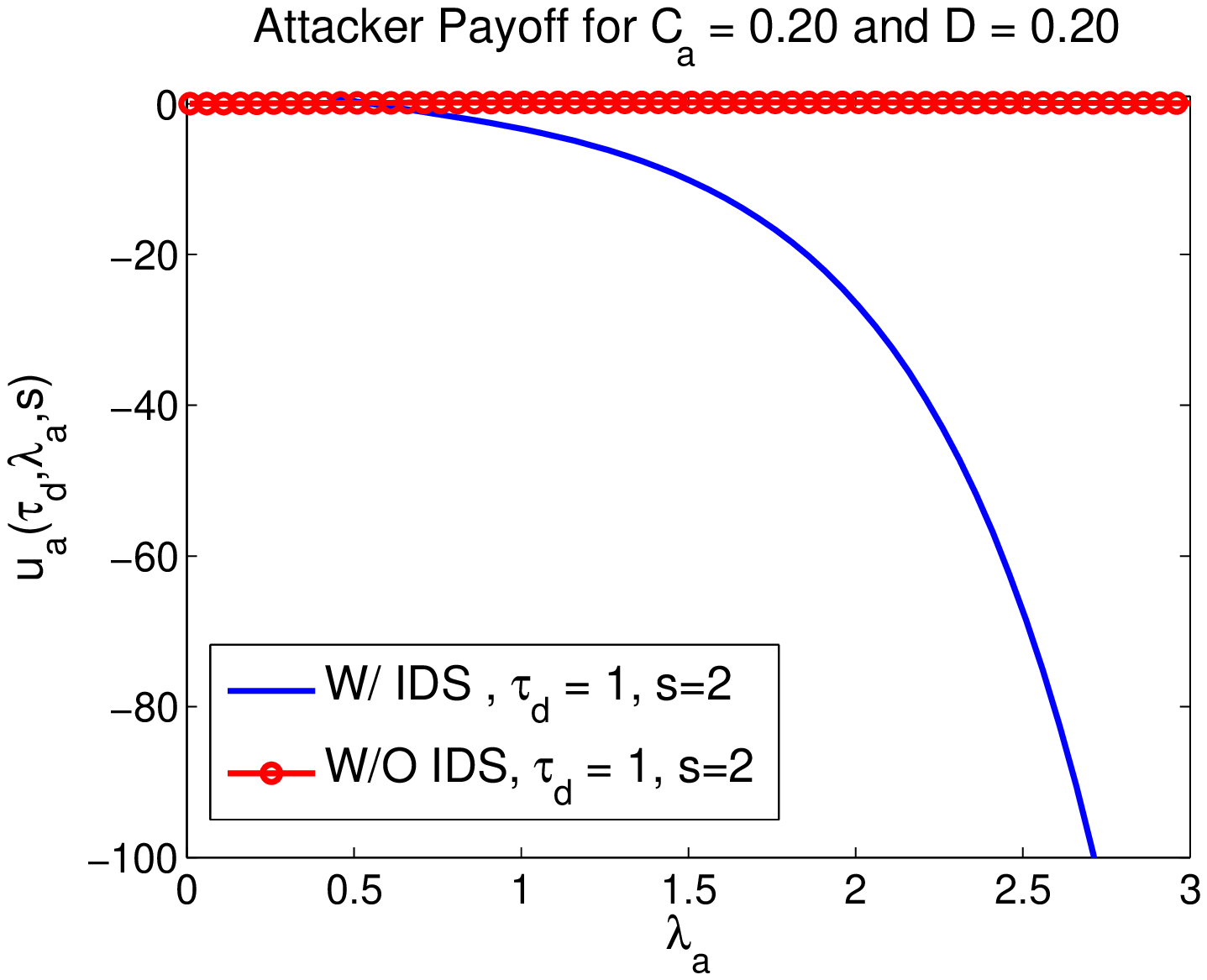}
\centering\caption{Attacker's payoff with and without IDS.}
\label{fig:extended}
\end{minipage}
\hspace{.05\linewidth}
\begin{minipage}{.45\textwidth}
  \centering
  \includegraphics[width=2.5 in]{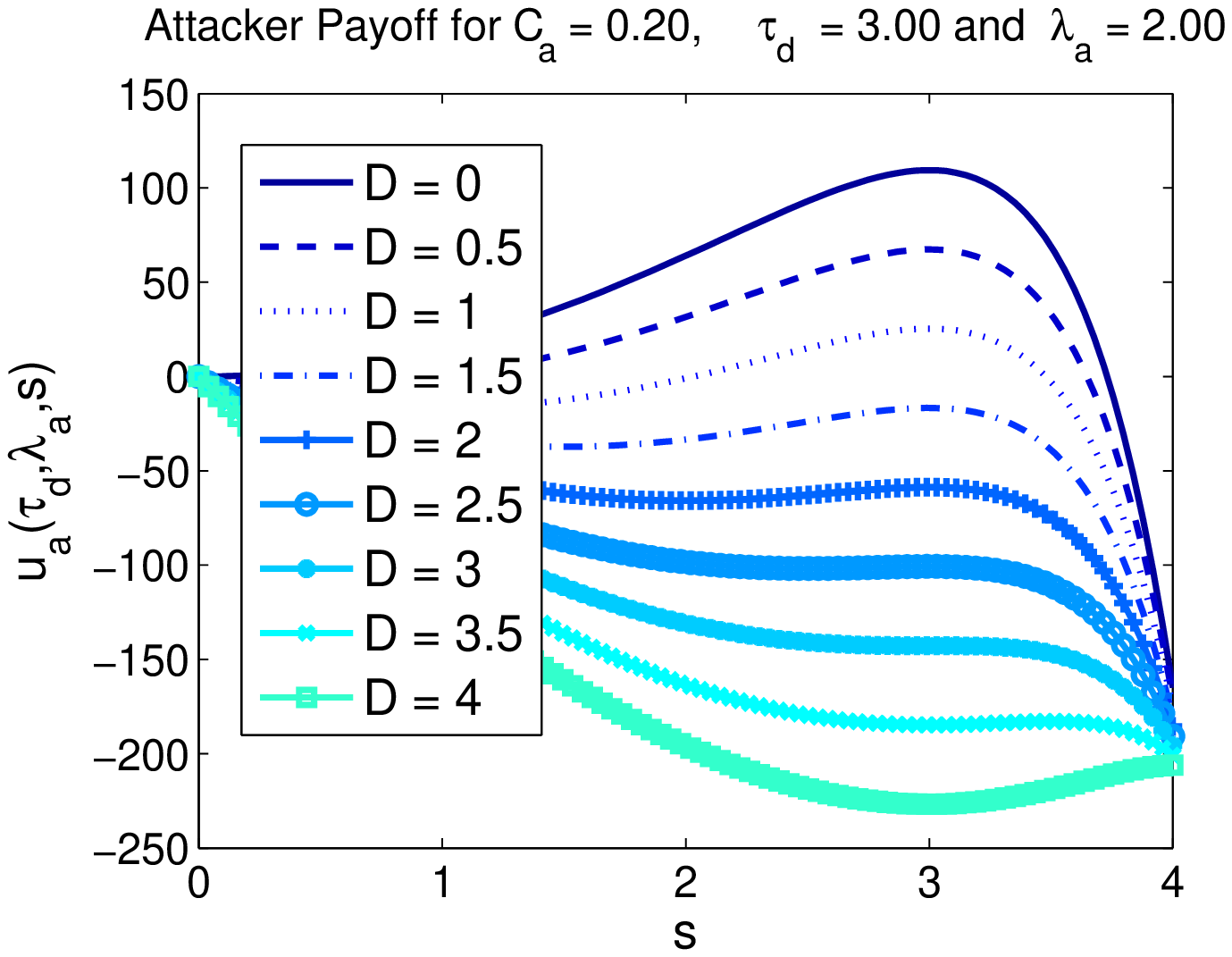}
 \centering\caption{Attacker's reward vs stopping time $s$ at different costs of detection.} \label{fig:extended_D}
\end{minipage}
\end{figure}

In Fig \ref{fig:extended}, we compare the attacker's payoff with and without an IDS in place based on the analysis in Section \ref{sec:extended}. In this experiment, we set $D = 0.2$, $C_a =0.2$ and $s = T/2$. It is clear that the IDS drastically reduces the attacker's reward. 
Fig. \ref{fig:extended_D} illustrates the attacker's payoff as function of the attack stopping time $s$ for different values of the detection cost $D$.  Obviously, the best time to stop the attack depends on the detection cost $D$. As $D$ increases, the attacker's payoff decreases and the optimal stopping time $s$ (corresponding to the highest payoff) is shown to decrease. At a certain point, the attacker is forced to stop as soon as she collocates to evade a high penalty if detected.  

\section{Conclusion}\label{sec:conc}
In this paper, We developed a moving target defense framework for the virtual machines migration timing problem. 
Live migration of virtual machines between different physical nodes is studied in a game-theoretic framework to defend multi-tenant clouds against side channel attacks launched by malicious users who are co-residing on the same physical node. We characterized best strategies for the players and established Nash equilibria existence conditions. We also considered an extended system model in which the Cloud is equipped with an intrusion detection system. We also verified our theoretical results numerically for different settings of the game. The theoretical and numerical analyses provided characterize the performance of the migration defense approach against collocation attacks.

\section*{Appendix A}

First we show that $u_a$ is strictly concave in $\lambda_a$. If $f_a(\tau_a; \lambda_a)$ is  {strictly} concave in $\lambda_a$, then 
\begin{equation}
f_a(\tau_a;\alpha \lambda_{a_1} + (1-\alpha)  \lambda_{a_2} ) > \alpha f_a(\tau_a;\lambda_{a_1} ) +  (1-\alpha)f_a(\tau_a;\lambda_{a_2} ).
\end{equation}
Hence,
\begin{equation}
\begin{split}
& \int_0^{\tau_d} G(\tau_a,\tau_d) f_a(\tau_a;\alpha \lambda_{a_1} + (1-\alpha)  \lambda_{a_2} ) d\tau_a > \\& \alpha  \int_0^{\tau_d} G(\tau_a,\tau_d) f_a(\tau_a;\lambda_{a_1} ) d\tau_a + (1-\alpha)\int_0^{\tau_d} G(\tau_a,\tau_d)  f_a(\tau_a;\lambda_{a_2}) d\tau_a.
\end{split}
\end{equation}
Therefore, it follows that
\begin{equation}
u_a(\tau_d, \alpha \lambda_{a_1} + (1-\alpha)  \lambda_{a_2} )  > \alpha u_a(\tau_d,  \lambda_{a_1} ) + (1-\alpha) u_a(\tau_d, \lambda_{a_2} ).
\end{equation}
Thus, $u_a$ is strictly concave in $\lambda_a$. To complete the proof, we further show that $u_d$ is strictly concave in $\tau_d$ if $\frac{G}{\tau_d}$ is convex in $\tau_d$. To this end, we show that if $\frac{G}{\tau_d}$ is strictly in $\tau_d$ then $
\frac{\partial^2 u_d}{\partial \tau_d^2} < 0$. 
%
Taking the derivative of $u_d$ in (\ref{eqn:defender_cond_rew}) with respect to $
\tau_d$
\begin{equation}
\frac{\partial u_d}{\partial \tau_d} = \frac{-G(\tau_d,\tau_d)}{\tau_d}f_a(\tau_d;\lambda_a) 
+ 
\int_0^{\tau_d} f_a( \tau_a;\lambda_{a } ) \cdot \frac{\partial}{\partial \tau_d}\left [ \frac{-G}{\tau_d}   \right ] d\tau_a + \frac{C_d}{\tau_d^2}.
\end{equation}
Since $G(\tau_d,\tau_d) = 0 $, the above derivative reduces to
\begin{equation}
\frac{\partial u_d}{\partial \tau_d} = \int_0^{\tau_d} f_a( \tau_a;\lambda_{a } ) \cdot \frac{\partial}{\partial \tau_d}\left [ \frac{-G}{\tau_d}   \right ] d\tau_a + \frac{C_d}{\tau_d^2}.
\end{equation}
Accordingly, the second derivative can be computed as
\begin{equation}
\frac{\partial^2 u_d}{\partial \tau_d^2} =  \int_0^{\tau_d} f_a( \tau_a;\lambda_{a } ) \cdot \frac{\partial^2}{\partial \tau_d^2}\left [ \frac{-G}{\tau_d}   \right ] d\tau_a - \frac{2C_d}{\tau_d^3}.
\end{equation}
\noindent
If $\frac{G}{\tau_d}$ is convex in $\tau_d$, then $- \frac{\partial^2}{\partial \tau_d^2}\left [ \frac{G}{\tau_d}   \right ] \leq 0$. Hence, 
$\frac{\partial^2 u_d}{\partial \tau_d^2} < 0 $, 
which completes the proof. 

\section*{Appendix B}

\subsection{\textbf{Proof of theorem \ref{theorem2} }}
\begin{proof}

From Theorem \ref{theorem1}, it suffices to show that if the inequality in the statement of the theorem is satisfied, i.e., 
$ 1-\lambda_a C_d < (1+\lambda_a \tau_d + \frac{\lambda^2_a \tau_d^2}{2})e^{-\lambda_a \tau_d}, \forall (\tau_d, \lambda_a) \in \mathcal{A}_d \times \mathcal{A}_a$, then
$u_a$ is strictly concave in $\lambda_a$ for every $\tau_d\in \mathcal{A}_d$ and $u_d$ is strictly concave in $ \tau_d $ for every $ \lambda_a  \in \mathcal{A}_a $. 

First, we show that $u_a$ is strictly concave in $\lambda_a$  for every $\tau_d \in \mathcal{A}_d$ with no restrictions. We compute the second derivative 
%
\begin{equation}\label{eqn:dua2_dlamda2}
\frac{\partial^2 u_a}{\partial \lambda_a^2} = \frac{e^{-\lambda_a \tau_d} \left ( \lambda_a^2 \tau_d^2 + 2 \lambda_a \tau_d - 2 e^{\lambda_a \tau_d} +2 \right )}{\lambda_a^3 \tau_d}.
\end{equation}
By Taylor's expansion of $e^x$ for $x > 0$, 
$ e^x > \frac{x^2}{2} + x +1$. Hence, 
%
\begin{equation}
\left (\lambda^2_a \tau^2_d +2\lambda_a \tau_d +2 \right )e^{-\lambda_a \tau_d} < 2\:.
\end{equation}
It follows that
%
\begin{equation}\label{eqn:dua2_dlamda2_inequality}
\left ( \frac{\lambda^2_a \tau^2_d}{\lambda^3_a \tau_d} + \frac{2\lambda_a \tau_d}{\lambda^3_a \tau_d} + \frac{2}{\lambda^3_a \tau_d} \right )e^{-\lambda_a \tau_d} - \frac{2}{\lambda^3_a \tau_d} < 0.
\end{equation}
From (\ref{eqn:dua2_dlamda2_inequality}) and (\ref{eqn:dua2_dlamda2}), the second derivative is negative, implying that the payoff $u_a$ is concave in $\lambda_a$ for all $\tau_d \in \mathcal{A}_d$. 
 
We now show that $u_d$ is strictly concave in $\tau_d$ under the condition in the statement of Theorem \ref{theorem2}. Computing the second derivative, 
\begin{equation}\label{eqn:dud2_dtau2}
\frac{\partial^2 u_d}{\partial \tau_d^2} = \frac{2\left ( 1-\lambda_a \tau_d-e^{-\lambda_a \tau_d} -\lambda_a C_d \right )}{\lambda_a \tau_d^3} + \frac{2\left ( 1-e^{-\lambda_a \tau_d} \right )}{\tau_d^2} - \frac{\lambda_a e^{-\lambda_a \tau_d}}{\tau_d}.
\end{equation}
If $1- \lambda_a C_d < \left(\frac{\lambda^2_a \tau^2_d}{2} + \lambda_a \tau_d +  1 \right )e^{-\lambda_a \tau_d} $, then 
\begin{equation}
1-e^{-\lambda_a \tau_d} - \lambda_a \tau_d \left (1 - e^{-\lambda_a \tau_d} \right) < \lambda_a \tau_d +\lambda_a C_d + \frac{\lambda^2_a \tau^2_d}{2} e^{-\lambda_a \tau_d}.
\end{equation}
Rearranging the terms of the above inequality and dividing by $\lambda_a \tau_d^3$, 
\begin{equation}
\frac{1-e^{-\lambda_a \tau_d}}{\lambda_a \tau_d^3}  + \frac{1-e^{-\lambda_a \tau_d}}{  \tau_d^2} - \frac{1}{ \tau_d^2} - \frac{C_d}{\tau_d^3} - \frac{\lambda_a e^{-\lambda_a \tau_d}}{2 \tau_d} < 0.
\end{equation}
Therefore, 
\begin{equation}\label{eqn:u_d_inequality_2}
\frac{2\left ( 1-\lambda_a \tau_d-e^{-\lambda_a \tau_d} -\lambda_a C_d \right )}{\lambda_a \tau_d^3} + \frac{2\left ( 1-e^{-\lambda_a \tau_d} \right )}{\tau_d^2} - \frac{\lambda_a e^{-\lambda_a \tau_d}}{\tau_d} < 0.
\end{equation}
By comparing (\ref{eqn:u_d_inequality_2}) and (\ref{eqn:dud2_dtau2}), we see that $\frac{\partial^2 u_d}{\partial \tau_d^2} < 0 $. Hence, $u_d$ is strictly concave in $\tau_d$.  
\end{proof}
\subsection{\textbf{ Proof of Theorem \ref{theorem5}}}

\begin{proof}

Given the reward function $G(\tau_d, \tau_a)$ in (\ref{eqn:linear_g_of_t}) and the distribution of $\tau_a$ in (\ref{eqn:f_a}), the attacker's expected payoff function can be expressed as in (\ref{eqn:u_a}). If the lower bound on $C_a$ in the statement of the theorem is satisfied, i.e., 
\[
C_a > \frac{1-(1+\lambda_{\max}\tau_d)e^{-\lambda_{\max}\tau_d}}{\lambda_{\min}^2}, 
\]
\noindent 
then,
\[
C_a > \frac{1-(1+\lambda_{a}\tau_d)e^{-\lambda_{a}\tau_d}}{\lambda_{a}^2}, ~~ \forall \lambda_a \in \mathcal{A}_a
\]
since the function in the numerator of the RHS of the inequality is monotonically increasing in $\lambda_a$ for $\lambda_a, \tau_d \geq 0$. Hence, 
\[
1-C_a \lambda_a^2 - (\lambda_a \tau_d +1) e^{-\lambda_a \tau_d} < 0\:.
\]
Dividing both sides by $\lambda_a^2 \tau_d$, 
\[
\frac{1-C_a \lambda_a^2 - (\lambda_a \tau_d +1) e^{-\lambda_a \tau_d}}{\lambda_a^2 \tau_d} < 0.
\]
\noindent
The left hand side of the above inequality is $\frac{\partial u_a}{\partial \lambda_a}$. Thus, $u_a$ is monotonically decreasing in $\lambda_a$, therefore, $\lambda_a^* = \lambda_{\min}$, which completes the proof.
\end{proof}

\subsection{\textbf{ Proof of Theorem \ref{theorem6}}}

\begin{proof}
Similar to the argument used in the proof of Theorem \ref{theorem5} above, if $C_d > \frac{1-(1+\lambda_{a}T)e^{-\lambda_{a}T}}{\lambda_a}$, then 
\[
C_d > \frac{1-(1+\lambda_{a}\tau_d)e^{-\lambda_{a}\tau_d}}{\lambda_a},~ \forall \tau_d < T
\]
since $ 1-(1+\lambda_{a}\tau_d)e^{-\lambda_{a}\tau_d}$ is monotonically increasing in $\tau_d \geq 0$.
Manipulating the above inequality and dividing by $\lambda_a\tau_d^2$, 
\[
0 < \frac{C_d\lambda_a-1+(1+\lambda_a\tau_d)e^{-\lambda_a\tau_d}}{\lambda_a\tau_d^2} = \frac{\partial u_d}{\partial \tau_d} \:.
\]
\noindent
Hence, $u_d$ is monotonically increasing in $\tau_d$, therefore, the best response $ \tau_d^*(\lambda_a) = T$. 
\end{proof}

\bibliographystyle{IEEEtran}

\bibliography{references}

\end{document}